\documentclass[12pt,preprint]{aastex}  

\usepackage{natbib}
\usepackage{epsfig}

\def\insitu{\hbox{$in~situ$}}

\def\epsCMa{\hbox{$\epsilon$ CMa}}

\def\temp{\hbox{$T$}}

\def\Rgd{\hbox{$R_{\rm gd}$}}
\def\PFe{\hbox{$P_{\rm Fe}$}}

\def\nts{\hbox{$n_{\rm TS}$}}
\def\nis{\hbox{$n_{\rm IS,Sun}$}}

\def\ebv{\hbox{$E(B-V)$}}

\def\NSII{\hbox{$N{\rm (S^+)}$}}
\def\NSiII{\hbox{$N{\rm (Si^+)}$}}
\def\NFeII{\hbox{$N{\rm (Fe^+)}$}}

\def\kms{\hbox{km s$^{\rm -1}$}}

\def\glong{\hbox{$l^{\rm II}$}}
\def\glat{\hbox{$b^{\rm II}$}}

\def\logNHH{\hbox{log$N$(H$_{\rm 2}$)}}
\def\NHII{\hbox{$N({\rm H^+})$}}
\def\NHI{\hbox{$N {\rm (H^o )}$}}
\def\NHIHII{\hbox{$N {\rm (H^o + H^+ )}$}}

\def\NH{\hbox{$N$(H)}}
\def\NHH{\hbox{$N$(H$_{\rm 2}$)}}
\def\NHeI{\hbox{$N({\rm He^o})$}}

\def\NOI{\hbox{$N({\rm O^o})$}}
\def\NNI{\hbox{$N({\rm N^o})$}}

\def\NCII{\hbox{$N({\rm C^+})$}}
\def\NCIIstar{\hbox{$N({\rm C^{+ *}})$}}
\def\NMgI{\hbox{$N({\rm Mg^{o}})$}}
\def\NMgII{\hbox{$N({\rm Mg^{+}})$}}

\def\logNHH2{\hbox{log$N$(H$^o$+2H$_2$)}}
\def\NHH2{\hbox{$N$(H$^o$+2H$_2$)}}
\def\nHH2{\hbox{$n$(H$^o$+2H$_2$)}}

\def\nHH{\hbox{$n({\rm H_2 }$)}}
\def\nHI{\hbox{$n({\rm H^o})$}}
\def\nH{\hbox{$n_{\rm H}$}}

\def\nHeI{\hbox{$n({\rm He^o})$}}

\def\NeI{\hbox{${\rm Ne^o}$}}
\def\NeII{\hbox{${\rm Ne^+}$}}
\def\ArI{\hbox{${\rm Ar^o}$}}

\def\Htot{\hbox{H$_{\rm tot}$}}
\def\HH2{\hbox{H$_{\rm H^o + 2H_2}$}}
\def\Xtot{\hbox{X$_{\rm tot}$}}
\def\XI{\hbox{${\rm X^o}$}}
\def\Xn{\hbox{X$^{\rm n}$}}

\def\n{\hbox{$n$}}

\def\cmtwo{\hbox{cm$^{-2}$}}

\def\cc{\hbox{cm$^{-3}$}}

\def\deeg{\hbox{$^{\rm o}$}}

\def\ne{\hbox{$n {\rm (e^- )}$}}
\def\HI{\hbox{${\rm H^o}$}}
\def\CII{\hbox{${\rm C^+}$}}
\def\CIIstar{\hbox{${\rm C^{+ *}}$}}
\def\NI{\hbox{${\rm N^o}$}}
\def\NII{\hbox{${\rm N^+}$}}
\def\OI{\hbox{${\rm O^o}$}}

\def\MgI{\hbox{${\rm Mg^o}$}}
\def\MgII{\hbox{${\rm Mg^+}$}}
\def\SII{\hbox{${\rm S^+}$}}
\def\HH{\hbox{${\rm H_2}$}}
\def\HII{\hbox{${\rm H^+}$}}

\def\SiII{\hbox{Si$^+$}}

\def\HeI{\hbox{He$^{\rm o }$}}
\def\HeII{\hbox{He$^{\rm + }$}}

\def\FeII{\hbox{Fe$^+$}}

\def\kms{\hbox{km s$^{-1}$}}
\def\cmtwo{\hbox{cm$^{-2}$}}

\def\22V{\hbox{d$V^{\rm 101}_{\rm s}$}}

\shorttitle{Chemical Composition of CLIC}
\shortauthors{Frisch \& Slavin}

\begin{document}
\title{The Chemical Composition and Gas-to-Dust Mass Ratio of Nearby
Interstellar Matter}

\author{Priscilla C. Frisch} 
\affil{University of Chicago, Department of Astronomy and
Astrophysics, 5460 S. Ellis Avenue, Chicago, IL 60637}
\author{Jonathan D. Slavin}
\affil{Harvard-Smithsonian Center for Astrophysics, 60 Garden
Street, MS 34, Cambridge, MA 02138}

\begin{abstract}

We use recent results on interstellar gas towards nearby stars and
interstellar byproducts within the solar system to select among the
equilibrium radiative transfer models of the nearest interstellar
material presented in \citet{SlavinFrisch:2002}.  For the assumption
that O/H$\sim$400 PPM, Models 2 and 8 are found to yield good fits to
available data on interstellar material inside and outside of the
heliosphere, with the exception of the Ne abundance in the pickup ion
and anomalous cosmic ray populations.  For these models, the
interstellar medium (ISM) at the entry point to the heliosphere has
\nHI=0.202--0.208 \cc, \nHeI=0.0137--0.0152 \cc, and ionizations
$\chi$(H)=0.29--0.30, $\chi$(He)=0.47--0.51.  These best models
suggest the chemical composition of the nearby interstellar medium
(ISM) is $\sim$60--70\% subsolar if S is undepleted.  Both \HI\ and \HII\
need to be included when evaluating abundances of ions found in warm
diffuse clouds.  Models 2, 8 yield an H filtration factor $\sim0.46$.
Gas-to-dust mass ratios for the ISM towards $\epsilon$ CMa are \Rgd$ =
178-183$ for solar abundances of \citet[][]{Holweger:2001}, or \Rgd$ =
611-657$ for an interstellar abundance standard 70\% solar.  Direct
observations of dust grains in the solar system by Ulysses and Galileo
yield \Rgd$\simeq$115 for models 2, 8, supporting earlier results
\citep{Frischetal:1999}.  If the local ISM abundances are subsolar,
then gas and dust are decoupled over small spatial scales.  The
inferred variation in \Rgd\ over parsec length scales is consistent
with the fact that the ISM near the Sun is part of a dynamically
active cluster of cloudlets flowing away from the Sco-Cen Association.
Observations towards stars within $\sim500$ pc show that \Rgd\
correlates with the the percentage of the dust mass that is carried by
iron, suggesting that a Fe-rich grain core remains after gain
destruction.  Evidently large dust grains ($> 10 ^{-13}$ g) and small
dust grains ($< 10 ^{-13}$ g) are not well mixed over parsec-length
spatial scales in the ISM.  It also appears that very small
Fe-dominated dust grains have been destroyed in the ISM within several
parsecs of the Sun, since C appears to be essentially undepleted.
However if gas-dust coupling breaks down over the cloud lifetime, the
missing mass arguments applied here to determine \Rgd\ and dust grain
mineralogy are not appropriate.
\end{abstract}

\keywords{ISM: abundances --- ISM: clouds --- dust}

\section{Introduction}

In this paper we combine recent observations of interstellar material
inside \citep[][CSS02, GG03
respectively]{Witteetal:2003,CummingsStone:2002,GloecklerGeiss:2003},
and outside \citep{GryJenkins:2001} of the heliosphere with radiative
transfer models which predict the boundary conditions of the
heliosphere \citep[][Paper II]{SlavinFrisch:2002}, to evaluate the
boundary conditions of the heliosphere, chemical composition of the
ISM nearest the Sun, and fundamental interstellar dust properties.  The
precise chemical composition and physical properties of low column
density ($\leq$10$^{18.5}$ \cmtwo) diffuse interstellar clouds are
difficult to determine because of unobserved ionization states,
component blending in distant sightlines, and unknown grain
composition.  The interstellar cloud surrounding the solar system
provides a unique opportunity to constrain the reference abundance
pattern of diffuse interstellar clouds and the composition of the gas
and dust phases of the interstellar medium (ISM).  For this cloud {\it
only} we have observations over parsec length scales towards nearby
stars, as well as \emph{in~situ} data sampling the cloud properties at
the entry point of the heliosphere.  \citet{Seaton:1951} examined the
ionization equilibrium of interstellar gas towards $\chi^2$ Ori and
recognized that interstellar dust grains (ISDGs) must be included for
a full understanding of the chemical composition of the ISM.
\citet[][hereafter Paper I]{Frischetal:1999} found that the
gas-to-dust mass ratio (\Rgd) in the interstellar cloud surroundign
the solar system (or the Local Interstellar Cloud, LIC) falls in the range
$\sim100-600$ when calculated with the ``missing mass'' method, which
compares the observed gas-phase abundances with an assumed reference
chemical composition; the uncertainties reflect the poorly known ISM
reference metallicity, and absorption line uncertainties.

In the special case of the interstellar cloud surrounding the solar
system (the local interstellar cloud, or LIC), direct measurements of
interstellar dust grains within the solar system have been obtained by
the Ulysses and Galileo spacecraft, giving a second means for
calculating \Rgd\ (Paper I).  The availability of two relatively
direct determinations of \Rgd\ for the LIC makes this cloud a unique
laboratory for evaluating the chemical composition of our Galaxy in
general, and diffuse clouds in particular.

The LIC is a member of an ensemble of cloudlets (denoted the cluster
of local interstellar clouds, CLIC) flowing from an upstream direction
(after vector conversion into the local standard of rest, LSR) within
$\sim$5\deeg\ of the Galactic center and towards the Loop I supernova
remnant interior.  The CLIC ensemble moves through the LSR at
17.0$\pm$4.6 \kms, and the velocity dispersion represents macroscopic
turbulence \citep{Frischetal:2002}.
\footnote{This direction is based on Hipparcos results, which yield a
solar motion in the local standard of rest of $\sim13.4$ \kms\ towards
the Galactic coordinates $l=28$\deeg, $b=+32$\deeg\
\citep{DehnenBinney:1998}.  For reference, the standard solar motion
yields an LSR upstream direction of \glong$\sim$330\deeg\
\citep{Frischetal:2002}.}

Observations of nearby stars sample various sightlines through the
CLIC.  Studies of stars with low \NHI\ values, e.g. $\epsilon$ CMa
\citep[][]{GryJenkins:2001}, REJ 1032+532 \citep{Holbergetal:1999} and
$\mu$ Col \citep{HowkSavageFabian:1999}, show that both neutral and
ionized gas are required for accurate abundance calculations of ions
formed in partially ionized clouds.  Data-constrained radiative
transfer models have been used to obtain both the chemical composition
and ionization levels of the ISM within $\sim3$ pc \citep[][hereafter
Paper II]{SlavinFrisch:2002}.  Such models yield gas phase abundances
referenced to \HI+\HII, which is important when determining abundances
for elements with first ionization potential (FIP) less than 13.6 eV,
as well as He, Ne, and Ar and gas-to-dust mass ratios.

The ionization corrections in the radiative transfer models (Paper II) are
imperative for low FIP ions in sightlines where comparable
amounts of \HI\ and \HII\ exist.  Clouds with \NHI$\lesssim10^{18.3}$
\cmtwo\ will be partially ionized, since for \HI, $\tau$(13.6 eV)=6 for
\NHI$ \sim10^{18}$ \cmtwo.  For the partially ionized gas within
$\sim3$ pc ($\chi{\rm (H)}  >$20\%, Paper II) both \HI\ and \HII\ must be included for abundance
determinations.  In particular, S$^+$ exists in \HI\ and \HII\ gas. Also, Fe,
Mg, Si and other refractories are significantly depleted so that neglecting
\HII\ in abundance estimates for partially ionized low density clouds will yield
both incorrect grain composition and inaccurate values for \Rgd.  This effect
is shown by the poor correlation between \FeII\ and \HI\ for low column
densities \citep[$\lesssim 10^{18}$ \cmtwo, Paper I,][]{WakkerMathis:2000},
versus the good correlation between \FeII\ and \MgII\ at the lowest column
densities (\NMgII$\lesssim 10^{12.5}$ \cmtwo, Paper I).

Recent data on the global properties of diffuse clouds, and
interstellar byproducts inside of the solar system (pickup
ions,\footnote{Pickup ions are formed from the ionization of
interstellar neutrals in the solar system, and subsequent Lorentz
force coupling of these ions to the solar wind.} PUI, neutral helium,
and anomalous cosmic ray,\footnote{Anomalous cosmic rays are formed
when convected pickup ions are accelerated to energies $<300$ MeV in
the termination shock regions of the solar wind.} ACR) are used here
to select the models best matching the combination of data on the ISM
inside and outside (but nearby), the heliosphere.  Paper II contains a
unique approach, in that both \insitu\ data and observations towards
$\epsilon$ CMa are used to constrain the nearby ISM.  Recent FUSE and
STIS results showing a relatively constant O/H ratio in the general
diffuse ISM provide additional constraints on the set of viable models
from among those presented in Paper II (\S \ref{sec:ism}).  The
improved \insitu\ data then yield two best models which are consistent
with most of the data (\S \ref{sec:insitu}).  The implications of
these models for the chemical composition of the local ISM (LISM),
dust grain composition, the gas-to-dust mass ratio at the entry point
to the heliosphere (\S \ref{sec:rgd}), and filtration factors are also
discussed.  Evidence for a nearby interstellar magnetic field, which
is required to maintain pressure balance with the hot gas of the Local Bubble, 
is discussed briefly (\S \ref{sec:B}).  A range of supplementary data are presented in the
Appendix, including the interstellar ionization levels predicted for
Models 2, 8, and 18.

\section{Best Model for Interstellar Gas within 3 pc  \label{sec:comp}}

Paper II presents 25 radiative transfer models of the ISM within
$\sim3$ pc, with input variables including the neutral column density
(\NHI) to the CLIC surface towards $\epsilon$ CMa (since it is poorly
known), cloud volume density (\nH), and interstellar magnetic field
strength (which determined the interface pressure).  (For convenience,
many Paper II results are plotted in the Appendix by model number.)
In this section we will show that Models 2 and 8 yield the best
agreement to the combined LISM data towards $\epsilon$ CMa and
observations of the ISM products within the solar system.  The
$\epsilon$ CMa sightline was chosen for completeness of data
\citep[][\S \ref{app:cma}]{GryJenkins:2001}, and because this star
dominates the near extreme ultraviolet (EUV) radiation field
\citep{Vallerga:1996}.  The models were forced to produce agreement
with the observed column densities of \CIIstar, \NI, \OI, \MgII,
\SiII, \SII, and \FeII\ towards \epsCMa, with no additional
assumptions about elemental abundances.  The combination of the
observed \NOI\ and assumed \NHI\ presets the O abundance, since \OI\
and \HI\ ionization are tightly coupled by charge exchange
\citep{FieldSteigman:1971}.  Both ionization levels and abundances are
predicted by the models for H, He, C, N, Mg, Si, S and Fe, and all
predicted abundances were referenced to \NHIHII.  For Ne and Ar, the
abundances are assumed to be solar abundances (see Section 2.3.1), and ionization levels
are predicted by the models.  These models were constrained by the
interstellar column densities of clouds within 3 pc towards \epsCMa,
and interstellar cloud properties at the solar location inferred both
from observations of the ISM and ISM products inside of the
heliosphere.\footnote{For a review of heliosphere properties see
\citet{Zank:1999}.}  The \insitu\ data used in Paper II include
direct observations of interstellar \HeI\ inside of the solar system, and
observations of pickup ions formed by the
interaction of interstellar neutrals with the solar wind.  In this
paper, we also use anomalous cosmic ray Ar data
\citep{CummingsStone:2002}, produced by accelerated pickup ions, to
select among models.  In Paper II, we found that Models 11, 17, and 18
provided reasonable agreement with the data.  Here we add information
about O/H in the disk ISM, and use new data on interstellar He
\citep[Ulysses data][]{Witteetal:2003}, pickup ions (GG03), and
anomalous cosmic rays (CSS02).  We will conclude below that Models 2
and 8 are the best matches to the available data from among the
initial set of models presented in Paper II.

\subsection{Uncertainties in the ISM Reference Abundance Pattern
\label{sec:solar}}

Dust grains are a poorly understood reservoir of material that
effectively mask the chemical composition of the ISM.  Prior knowledge
of the ISM reference abundance pattern in principle allows the use of
volatile trace elements (e.g. S, O, N) as a hydrogen column density
proxy.  One possible abundance standard for the ISM is the solar system
abundance pattern \citep[][GS98]{GrevesseSauval:1998}, although recent
studies of the solar photosphere \citep[][H01]{Holweger:2001} and
solar granulation \citep[][Pr02]{PrietoAsplund:2002} make these values
less certain.  Interstellar reference abundances are also poorly
understood, although the subsolar B-star abundances have been
suggested as a template abundance pattern for the ISM
\citep[][]{Yorketal:1983B,SavageSembach:1996,SnowWitt:1996,SofiaMeyer:2001}.
The assumption of a solar reference abundance for the ISM can not be
tested for refractory elements such as Fe, which are heavily depleted
in disk and halo clouds.  The reference abundances for C, N, O, S, and
other elements in the ISM have been discussed extensively, including
the possibility that either B-star or another subsolar abundance
pattern is more appropriate as a standard for the ISM
\citep[][]{Yorketal:1983B,SavageSembach:1996,SnowWitt:1996,SofiaMeyer:2001}.
The lower solar O/H abundances found by H01 and Pr02 remove one
argument in favor of subsolar interstellar O abundances, which were
needed to keep the dust-entrained O at reasonable levels
\citep{SofiaMeyer:2001,Cartledgeetal:2001}.  If gas and dust are
poorly mixed over subparsec distance scales (\S \ref{sec:rgd}), and
gas and large dust grains (radius $>$1 $\mu$m) are never well mixed
\citep{GruenLandgraf:2000}, then the assumption that the ISM chemical
composition is the sum of the gas and dust components will be
incorrect.  Our conclusion below, that the two best models favor
$\sim$60--70\% subsolar abundances for the LISM (\S\ref{sec:abun}), is
derived independent of any assumptions about the dust composition,
and is consistent with interstellar Kr results showing abundances
$\sim$60\% solar (or meteoritic) values
\citep{CardelliMeyer:1997,Cartledgeetal:2001}.  As a noble element, Krypton
should not be depleted onto dust.  However our conclusion, that abundances 
in the nearby ISM are
likely subsolar, conflicts with findings of solar abundances in 
higher column density sightlines(\NH$>$10$^{20}$ \cmtwo) when H01 or Pr02
solar values are invoked
\citep{SofiaMeyer:2001}, so this problem is unresolved.

One possible H proxy is \SII, which is expected to be undepleted in
diffuse clouds and is formed in both \HI\ and \HII gas.  Towards
\epsCMa, \NSII $=1.35\times10^{13}$ \cmtwo\ (Appendix) giving log \NH$
=17.79$ \cmtwo\ for solar abundances ($\mathrm{S/H} = 22$ PPM, from
H01).  Models 1, 4, 7, 10, 13, 14, 16, 17, 22, and 25 yield log
\NHIHII\ predictions that are within 0.06 dex of this value (Table 4
of Paper II), but the predicted C/H for these same models (427--832
PPM) substantially exceeds the allowed solar C/H abundances of
245--391 (GS98, H01, Pr02, Table \ref{table1}).
Uncertainties of $\sim 25$\% on \NSII\ and 23\% on the solar S/H ratio allow
for agreement with several other models.  We find below that Models 2
and 8 (the two best models, \S 2.2, \S 2.3) suggest that
interstellar sulfur abundances in the $d<3$ pc ISM are subsolar.

Observations of diffuse clouds in disk sightlines show that O/H is
relatively constant towards stars with log \NHH2 $<21.5$ \cmtwo\
\citep{Cartledgeetal:2001,AndreHowketal:2003}.  Observations of the
\ion{O}{1} $\lambda$1356 \AA\ absorption line toward 13 stars within
$\sim$500 pc, with \NHH2 $ = 1-2\times10^{21}$ \cmtwo, and showing an
average spatial density $\langle n \rangle = 0.63$ \cc, yield O/H $
=347\pm16$ PPM \citep[][corrected to oscillator strength $f =
1.16\times10^{-6}$] {MeyerJuraCardelli:1998}.  Observations of 19
somewhat more distant stars (1--5 kpc) with column densities \NHH2
$\ge 10^{21}$ \cmtwo, and an average spatial density of $\langle n
\rangle=0.38$ \cc, yield O/H $ =408\pm14$ PPM
\citep{AndreHowketal:2003}.  Both studies find that O/H is relatively
constant over the sightlines sampled, although above log \NHH2 $ \sim
21.5$ \cmtwo\ O depletions appear to increase.  The four stars with
the largest O/H ratios in the Meyer et al. sample also give averages
$\langle$O/H$\rangle\sim$400 PPM and $\langle n \rangle \sim0.18$ \cc.
The Andre et al. star set samples disk stars with lower average volume
densities (since they are more distant) that are more similar to the
$\langle n \rangle \sim 0.3$ \cc\ found locally.  Thus we select O/H $
\sim$400 PPM as a suitable value for the ISM oxygen abundance close to
the Sun.  This O/H abundance range is also consistent with
observations of the nearby stars G191-B2B, WD0621-376, WD1634-573,
WD2211-495 \citep[to within 1$\sigma$
uncertainties,][]{Moosetal:2002}.

\subsection{Selecting Models Consistent with the Disk ISM O$^{\rm o}$/H$^{\rm o}$ 
 \label{sec:ism}}

Since the \HI\ column density towards \epsCMa\ is unknown, \NHI\ was
treated as a free parameter in Paper II.  In this section we will
require consistency between the LISM abundance and the general disk
ISM in order to narrow the selection of viable models.  We assume O/H
$ \sim 400$ PPM to obtain \NHI $ = 6.5\times10^{17}$ \cmtwo\ for the
\epsCMa\ sightline, since \NOI $ = 2.6^{+0.8}_{-0.5}\times10^{14}$
\cmtwo\ \citep[][and Appendix]{GryJenkins:2001}.  For comparison, EUVE
broadband observations found \NHI$ = 9.5\pm2.5\times10^{17}$\cmtwo\
\citep{Vallerga:1996} towards $\epsilon$ CMa, and model atmosphere
calculations are consistent with an interstellar column of \NHI$ =
5\times10^{17}$ \citep{Aufdenberg:1998}.  Hence, the most viable
models from Paper II are Models 2, 5, 8, 11, 18, 20 and 23, which
predict O/H=380--389 PPM.  Note this O abundance is 72\% (H01), or
82\% (Pr02) of the solar abundances (Table \ref{table1}, Appendix).  In
the following sections (\ref{sec:ism} and \ref{sec:insitu}) we argue
that Models 2 and 18 are the best matches to the combined interstellar
and \insitu\ data.  Note that none of our models have O/H$\sim$265
PPM, as seen towards higher column density interstellar clouds
\cite{Cartledgeetal:2001}.

Nitrogen shows no systematic abundance or depletion variations with
variable column density
\citep{Yorketal:1983B,MeyerCardelliSofia:1997}.  Towards $\epsilon$
CMa, \NNI/\NHI$ =41\pm2$ PPM (for \NHI$ = 6.5\times10^{17}$ \cmtwo),
and the viable models (Models 2, 5, 8, 11, 18, 20 and 23) have
predicted N abundances of $43.7-52.5$ PPM.  The difference between
\NNI/\NHI\ and N/H is because N and H ionization decouples in high
radiation fields. Nitrogen is dominated by charge exchange with H,
except for some photoionization at the cloud surface (e.g. Fig. 2,
Paper II).  \citet{MeyerCardelliSofia:1997} find an interstellar
abundance of $75\pm4$ PPM, however this value seems too large for
diffuse clouds since FUSE observations towards low column density
sightlines ($< 10^{20}$ \cmtwo) show \NNI/\NHI$ = 35-60$
\citep[][]{Moosetal:2002}.  For example, \NI/\HI$ = 38 \pm 8$ PPM
towards HZ43 \citep[\NHI$\sim8.7\times10^{17}$,][]{Kruketal:2002};
towards Capella \NI/\HI$ = 42 \pm 2$ PPM
\citep{WoodRedfieldetal:2002}.  CLIC N abundances appear to be $\sim
51$\%--78\% of solar (Table \ref{table1}).  Models 2 and 8 find that
N/H$\sim50$, compared to H01 solar abundances of N/H$ =
85^{+25}_{-19}$, indicating interstellar N abundances that are
$\sim$60\% of solar values.

Sulfur is generally considered undepleted in the ISM
\citep{SavageSembach:1996} and \SII\ is the dominant state of S in
both warm neutral material and warm ionized material since FIP=10.4
eV.  Neglecting \HII\ yields erroneous S abundance values in partially
ionized gas such as the CLIC.  The restricted set of models gives a S
abundance range $13.5-15.5$ PPM, compared to the solar abundance
$22\pm5$ PPM (Table \ref{table1}).  Had \HII\ been ignored in this low
column density sightline, S would appear to have solar abundances of
21 PPM.  Low interstellar S abundances are also found in HST
observations of REJ1032+532, with a S abundance of 9.5 PPM
\citep[using $N$(\HI+\HII)$ \sim 8\times10^{18}$
\cmtwo,][]{Holbergetal:1999}.  However, the ISM towards $\mu$ Col
shows solar S abundances to within uncertainties, \citep[using log
$N$(\HI+\HII)$ \sim 19.9$ \cmtwo,][]{HowkSavageFabian:1999}.  Based on
the abundances predicted for Models 2, 5, 8, 11, 18, 20 and 23, S is
present at $\sim$50\%--90\% of solar in the CLIC, with Models 2 and 8
yielding 60\% solar abundances.

Nearly all C in warm diffuse clouds is singly ionized (IP=11.3 eV).
Carbon abundances for the viable models are in the range 263--275 PPM,
compared to solar abundances $331^{+49}_{-43}$ (GS98),
$391^{+110}_{-86}$ PPM (H01), or $245^{+24}_{-21}$ (Pr02).  Models 2
and 18, selected below as the best models, yield C/H=263 PPM, and 309
PPM which exceeds the solar abundances found by (Pr02).  The current
picture of the ISM does not include supersolar C abundances in the gas
phase (although poor dust-gas mixing combined with grain destruction
might produce supersolar C/H gas values in small cloudlets).  Thus we
adopt the H01 solar abundances which imply C/H abundances that are
67\%--79\% solar.  The mild C depletion may indicate the destruction
of small dust grains, or PAHs, which is consistent with evidence that
the CLIC has been shocked (Paper I).  FUSE values for low column
density sightlines are in the range \NCII/\NHI $ = 350-800$ PPM, with
large uncertainties, versus \NCII/\NHI $ =423\pm102$ PPM for the
\epsCMa\ sightline and \NHI $ = 6.5 \times 10^{17}$ (Appendix).
Including the fact that $\sim56$\% of the gas towards HZ43 is ionized
\citep{Kruketal:2002} reduces C/H to $\sim345$, which is closer to the
CLIC value.  Model 19, which has been suggested elsewhere as the best LIC model,
can be rejected since it predicts a C abundance (759 PPM)
substantially larger than the solar value.

The ratios \MgII/\MgI\ and \CII/\CIIstar\ provide diagnostics of the
cloud ionization.  Among the viable models, Models 2, 8, and 18 give
good agreement with \MgII/\MgI\ towards $\epsilon$ CMa (346$\pm$87),
and Model 11 is consistent (Table \ref{table2}).
The larger predicted ratios for \MgII/\MgI\ for Models 5, 20 and 23 are
inconsistent with the $\epsilon$ CMa data.  
The uncertainties on \NCII\ are large, and all of the viable models
are consistent with the observed \CII/\CIIstar\ ratio towards
$\epsilon$ CMa to within the uncertainties.  However, Models 2, 8, and
18 provide the best match.  Predicted abundances for these models are
compared to solar and the ISM reference abundances in Table
\ref{table3}.  Based on these ratios and O/H$\sim$400 PPM, Models 2,
8 and 18 provide the best match to the \epsCMa\ data.
The \MgII/\MgI\ and \CII/\CIIstar\ ratios for all models,
and the ionization fractions for Models 2, 8, and 18, are given in the Appendix.

\subsection{Constraining Models with \insitu\  Observations of the ISM 
\label{sec:insitu}}

\emph{In~situ} observations of the products of the solar wind
interaction with interstellar neutrals fed into the heliosphere by the
surrounding cloud yield a sample of the ISM at the entry point to the
heliosphere.  The LIC is a special case because it is the only
interstellar cloud with data from a single location (the entry point
to the heliosphere) rather than sightline averaged values.  However,
\emph{in~situ} observations of interstellar neutrals and the
byproducts of the ISM-heliosphere interaction introduce uncertainties
from the uncertain neutral-ion interactions in the heliosheath regions
between the bow shock (if it exists) and the solar wind termination
shock \citep{RipkenFahr:1983}.  These interactions have been
calculated
\citep{Izmodenovetal:1999b,CummingsStone:2002,MuellerZank:2002}.  The
filtration factor, $F_{\rm X}$ for an element X, is the ratio of the
densities of X$^{\rm o}$ inside and outside of the heliosphere (since
only the neutrals are able to penetrate the heliosphere).  Filtration
factor estimates are based on either a heliosphere model, whose
boundary conditions are set by the physical properties of the ISM at
the solar location, or a model of the ISM at the entry point to the
heliosphere (\S \ref{sec:filt}).  The need to decouple the calculated
filtration factor from any assumptions about the heliosphere model
helped to motivate this study.  The most recent \insitu\ data are used
below to further select among the viable models.  It is shown below
that the pickup ion data suggest Models 2 and 8 agree best with the
combined astronomical and \insitu\ data.

\subsubsection{Helium, Oxygen, Nitrogen, Neon, Argon}

Direct observations of interstellar \HeI\ in the solar system by
Ulysses yield a temperature of 6,300$\pm$340 K and density
0.014$\pm$0.002 \cc\ for He at the termination shock
\citep[][]{Witte:1996,Witteetal:2003}.  Observations of He in the
pickup ion population give \nHeI$ = 0.015\pm0.002$ \cc\ at the
termination shock \citep{GloecklerGeiss:2003}.  We adopt
\nHeI=0.0145$\pm$0.015 \cc, which is the overlap between these
independent measurements.  Helium has minimal filtration ($F_{\rm
He}=0.94-0.99$, Table \ref{table3}) in the heliosheath regions.
Correcting for \HeI\ filtration yields a LIC density at entry to the
heliosphere of \nHeI=0.013--0.017 \cmtwo, which encompasses the
predicted \nHeI\ values at for Models 2, 5, 8, 11, 18 and 23.  Models
2 and 8 predict \nHeI=0.015 and 0.014 \cc.

The pickup ion data yield \OI/\HeI$ = (3.66\pm0.67)\times10^{-3}$ for
interstellar neutrals at the upstream termination shock (GG03 and 
Table \ref{table3}).  The viable models yield predictions of \OI/\HeI$ = (5.00 -
5.76)\times10^{-3}$.  Since over 30\% of the \OI\ atoms may be removed by
charge exchange between \OI\ and interstellar protons in the heliosheath
regions, the
predictions are consistent with the data.  Comparison between the
models and pickup ion data, however, do appear to rule out the
augmentation of \OI\ in the heliosheath \citep[or a filtration factor
$>1$,][]{MuellerZank:2002}.  Models 2 and 8 yield $F_{\rm O}$=0.6 for
the O filtration factor (eq. 1).

The pickup ion data yield \NI/\HeI$ = (5.38\pm1.17)\times10^{-4}$ for
interstellar neutrals at the termination shock.  The viable models
predict \NI/\HeI$ = (5.11-6.12)\times10^{-4}$, and these values are
within the PUI uncertainties when filtration factors are included
\citep[$\sim0.76-0.96$][Table \ref{table3}]{MuellerZank:2002,
CummingsStone:2002}.  Models 2, 8, and 18 imply a N filtration factor
of $\sim 0.85$ (eq. 1).

Neon is observed in the PUI and ACR populations, and ACR Ar is
detected.  As noble elements, Ne and Ar should not deplete onto dust
grains, and therefore should indicate the ISM reference abundance.
Both \ArI\ and \NeI\ are sensitive to the EUV radiation field, and the
viable models predict \ArI/Ar=22--33\%, and \NeI/Ne=9--18\%.  In Paper
II Ne and Ar abundances were assumed at 123 PPM and 2.82 PPM
respectively.  However the solar Ar abundance in GS98 is
$2.51^{+0.37}_{-0.33}$ PPM, and B-star Ar abundances appear somewhat
larger \citep[$3.16^{+0.39}_{-0.34}$ PPM,][] {Holmgrenetal:1990}.  In this
paper we use the Holmgren Ar abundance, and correct the Paper II
predictions for this 12\% abundance increase.

Anomalous cosmic ray data yield \ArI/\HeI$ =
(1.50\pm0.67)\times10^{-5}$ at the termination shock
\citep{CummingsStone:2002,GloecklerGeiss:2001}.  Using Voyager data
and new calculations of filtration factors, Cummings et al.  selected
Model 18 as the best match to the anomalous cosmic ray data.  Below we
show that Models 2 and 8 both yield satisfactory fits to the
combination of \epsCMa\ and \emph{in~situ} data.  

Paper II noted the failure of the models to predict the pickup ion Ne
values for an assumed solar abundance of 123 PPM.  Neglecting Ne
filtration, Models 2 and 8 predict \NeI/\HeI\ values that are 41--60\%
of the pickup ion value \NeI/\HeI$ = (5.24\pm1.18)\times10^{-4}$ at
the termination shock (GG03, Table \ref{table3}).  The inclusion of Ne
filtration ($\lesssim$12\%) aggravates the difference between the
model results and PUI \NeI\ data.  Models 5 and 20 predict the most
\NeI, but yield poor agreement with the $\epsilon$ CMa data.  Reducing
the Ne discrepancy requires either reducing Ne ionization, or raising
the intrinsic interstellar abundance to $\sim$200 PPM.  Since
$\sim$12\% of the Ne is neutral, relatively small reductions in the
radiation field may raise \NeI/\HeI\ by the required amount.  However
the FIPs of \NeI\ (21.6 eV) and \HeI\ (24.6 eV) are similar, and the
\HeI\ ionization predictions are satisfactory (Table \ref{table3}).
An unexplored possibility is that charge exchange between \HeI\ (ionization
potential $\sim$24.6 eV) and \NeII\ in the heliosheath may amplify Ne
inside of the heliosphere.

\subsubsection{Hydrogen Pickup Ions and Filtration\label{sec:filt}}  

In Paper II we did not use observations of \HI\ or PUI H in the solar
system to constrain the models because of the poorly known but large
filtration factor from charge exchange between \HI\ and \HII\ in the
heliosheath \citep{RipkenFahr:1983,Izmodenovetal:1999}.  However, we
now use these data to show that Models 2 and 8 provide the most
consistent agreement with the PUI H data.

\begin{figure}[ht]
\vspace*{4.8in}
\includegraphics{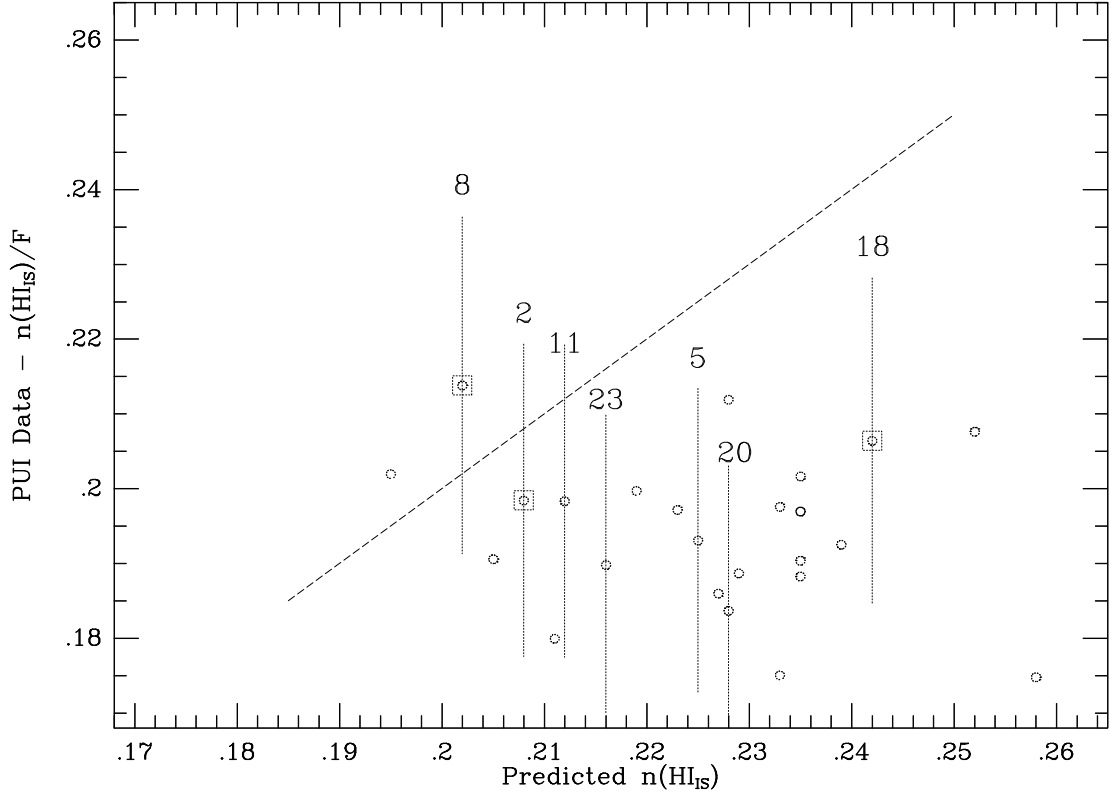}
\caption{
Comparison between \HI\ in the pickup ion population and model predictions for
\nHI\ for the 25 models.  The ordinate 
is the interstellar \nHI\ value that results after the pickup ion measurement
of \nHI\ at the termination shock is corrected for heliosheath filtration (see
text).  The abscissa is the predicted \nHI\
at the solar location.  The viable models show the uncertainties due to
the pickup ion data, and the three best models based on \MgII/\MgI\ and
\CII/\CIIstar\ ratios have boxes around the data points.  (The points for
Models 16 and 17, which differ only in the FUV radiation field, are
superimposed on each other.)
}
\label{fig1}
\end{figure}

Following \citet[][GG01]{GloecklerGeiss:2001}, assuming negligible He filtration,
the filtration factor for element X can be evaluated using: 
\begin{equation}
F_{\rm X} = \frac{\nts(X)/\nts(He)}{\nis(\XI)/\nis(\HeI)}.
\end{equation}
GG01 calculated the H filtration factor ($F_{\rm H}$) using pickup ion
data for the H and He densities at the termination shock (TS, \nts(H)
and \nts(He)), and assuming \nis(\HI)/\nis(\HeI)$ = 11.25$ for the
interstellar ratio at the entry point to the heliosphere.  The value
11.25 for \mbox{\nis(\HI)/\nis(\HeI)} is too low for the solar
location, although appropriate for sightline averaged
values.\footnote{ It seems unlikely that a new set of model parameters
with a lower initial assumed total volume density would yield a ratio
close to the 11.25 value assumed by GG03.  Comparisons between Models
19 and 25, which have the same input variables except for \n\ (which
is 0.273 \cc for Model 19 and 0.227 for Model 25) shows that a 27\%
decrease in the volume density yields a 3\% increase in
\nis(\HI)/\nis(\HeI) at the solar location.}  Paper II found sightline
averaged ratios of \NHI/\NHeI$ = 8.9-13.6$.  At the solar location,
\nis(\HI)/\nis(\HeI)$ =13.7-14.7$ since the \HI\ opacity exceeds the
\HeI\ opacity to cloud surface, yielding $F_{\rm H}$=0.45, 0.48 for
Models 2, 8.  Models 2, 5, 11, 18, 20, and 23 give H filtration
factors in the range 0.41--0.48.  The first row of Table \ref{table3}
gives the model predictions for interstellar \nHI\ at the entry point
to the heliosphere.  The second row gives the interstellar \nHI\ calculated with
a semi-empirical filtration factor applied to the pickup data \nHI\ at the
termination shock, assuming negligible He filtration.  
Each of the interstellar \nHI\ values predicted using model-corrected
filtration factors are within 15\% of the \nHI\ value at the solar
location actually predicted by the model (the first row of Table
\ref{table3}).  
Fig. \ref{fig1} shows interstellar \nHI$_{\rm is}$ derived
from the PUI value at the TS corrected for filtration
(ordinate, \nHI$_{\rm is}$=\nts(\HI)/$F_{\rm H}$))
plotted against the interstellar \nHI\ predicted by the models 
(column 10, Table 7 of Paper II).
The viable models points are numbered, while the points for Model 2, 8, and 18, which are favored by
the Mg and C data, are enclosed in boxes.  
Among the three best models favored by the $\epsilon$ CMa Mg and C
data (2, 8 and 18), Models 2 and 8 yield the best self-consistency between model-predicted
interstellar \HI\ volume densities, and model-corrected interstellar densities
derived from the PUI data.  Based on this comparison, we select Models 2 and 8
as the best overall matches to the available data on the LISM within
$\sim$3 pc, and conclude that this self-consistency validates the plausibility
of these results.

\subsubsection{Temperature}

The Ulysses observations of interstellar \HeI\ yield a LIC cloud temperature
temperature $6,300\pm 340$ K, compared to 8,200--8,500 K predicted by Models 2 and 8.
Model 25 has the lowest predicted temperature, \temp$ =
5,120$ K, but was excluded because of unrealistically large C abundances,
and disagreement with $\epsilon$ CMa \MgII/\MgI\ and \CII/\CIIstar\ ratios.
Typical radiative recombination rates in the ISM are on the order of
\temp$^{-2/3}$, so a predicted temperature that is too high by 1,500 K results
in a recombination rate that is too small by $\sim$14\%. 
\citet{Heiles:2001} concluded from comparisons of 21 cm emission and
absorption that over 47\% of warm neutral gas is in a thermally unstable
region corresponding to 500--5000 K, a range near the LIC temperature.  
However, Models 2 and 8 give H ionization $\chi {\rm (H)} \sim$0.30, 
and \nHI$\sim$0.20 \cc, and this partially ionized cloud type is in a regime not well studied theoretically.

\section{Discussion \label{sec:disc}}

The properties Models 2, 8 and 18 are summarized in Table \ref{table4}.  
The implications of these models are now discussed.

\subsection{The Extent of the ISM towards the Nearest Stars \label{sec:fill}}

Models 2 and 8 give \nHI$ =0.208$ \cc\ and 0.202 \cc\ for $d<$3 pc
ISM, based on the LISM towards $\epsilon$ CMa.  If all ISM close to
the Sun shares the characteristics of the LIC, then the thickness of
the clouds towards nearby stars can be estimated.  Towards $\alpha$
Cen (1.3 pc, the nearest star) log \NHI$ =17.60-18.02$ \cmtwo\
\citep[the lower limit corresponds to D/H$ =
1.5\times10^{-5}$,][]{LinskyWood:1996}.  Thus for $<$\nHI $> \sim$0.20
\cc\ (Models 2, 8), LIC-like gas would fill 60\%--100\% of the
sightline.  In the downstream direction, \NHI$ =
6.5^{+2.5}_{-2.0}\times10^{17}$ \cmtwo\ towards Sirius AB (2.7 pc)
\citep[e.g.][]{Hebrardetal:1999}.  In this case, LIC-like gas will
fill $\sim$37\% of the sightline.  Towards Procyon (3.5 pc) in the
downstream direction, log \NHI$ =18.06$ \cmtwo\ \citep[for both
components,][]{Linskyetal:1995}, and a LIC-like cloud fills $\sim$50\%
of the sightline.  Towards Capella (12.5 pc), the LIC gas extends
$\sim$2.6 pc, filling $\sim$21\% of the sightline (Table
\ref{table1}).  Towards $\lambda$ Sco (216 pc) where \NHI=10$^{19.23}$
\cmtwo\ \citep{York:1983}, LIC-like gas extends $\sim$25 pc.
The star $\lambda$ Sco samples a sightline near the upstream CLIC direction in
the LSR \citep[][]{Frischetal:2002,Frisch:1995}.  From this limited
data sample, Models 2 and 8 suggest that LIC-like gas fills less than
50\% of the sightline in downstream directions even for the nearest
stars.  For the upstream direction \HI\ data is sparse, and in
addition the dispersion in cloudlet velocities and the large extent of
the CLIC gas indicate that the LIC cloudlet model may not be
applicable.

\subsection{LISM Reference Abundances and Depletions\label{sec:abun}}

The chemical composition for nearby gas given by Models 2 and 8 (which
provide the best overall agreement with the ISM in the \epsCMa\
direction and in the heliosphere, \S \ref{sec:ism}, \S
\ref{sec:insitu}) is summarized in Table \ref{table1}.  
The gas phase interstellar abundances predicted by Model 8 (Model 2)
assuming no depletions are: C---275 (263); N---53 (50); by assumption O---380
(380); and S---13.2 (13.2).  For C, N, O, and S, respectively,
these values are 70\% (67\%),
62\% (59\%), 70\%, and 60\% of the H01 solar values.

Sulfur is not expected to be depleted onto dust grains in the ISM
\citep{SavageSembach:1996}.  The low S abundances found here suggest
the correct reference abundance for the LISM is $\sim$60--70\% of the H01
solar abundances.  An ISM reference abundance pattern of $\sim$60\%
solar is also consistent with Kr abundances that are $\sim$60\% solar
\citep{CardelliMeyer:1997}.

The depletion of C onto dust grains within $\sim3$ pc is not required by these
results, providing the abundances of the alpha-particle elements C, O, S, Kr are all
subsolar by similar amounts (60--70\%).  Since C is a common constituent of
dust grains, an interstellar gas phase abundance of $\sim$70\ solar leaves
minimal carbon for interstellar dust grains.
However, \citet{SofiaMeyer:2001} note that different elements in the ISM may have
different origins.  The Si, Fe, and Mg abundances are consistent
with grain destruction in interstellar shocks (Paper I), so the
small carbon-bearing grains (possibly PAH's) may also have been partially destroyed in the
nearest interstellar gas.  

The PUI and ACR populations are consistent with subsolar O and N 
abundances in the parent interstellar population.  Models 2 and 8
match the H, He, N, O, and Ar PUI and ACR data, for reasonable
filtration values.  Note, however, that the uncertainties on the O, Ar
and N filtration factors are comparable to the amount of subsolar LIC
metallicity.  The PUI and ACR Ne data are difficult to interprete in
terms of the models.  Filtration models predict a Ne loss of
$\sim$12\% in the heliosheath ($F_\mathrm{Ne} \sim0.88$, Table
\ref{table3}), implying either supersolar Ne abundances in the LISM,
or that the EUV radiation field is incorrect at the ionization
potential of Ne (21.6 eV).  Suprasolar Ne abundances are unlikely if
O, N, and S abundances are subsolar.  Models 2 and 8 show that Ne is
12\% neutral at the solar location, so small uncertainties in \NeI\
ionization levels, which are on the the EUV radiation field, will
yield large variations in \NeI/Ne.  This Ne discrepancy may reflect
the uncertainty in the diffuse EUV radiation field, which has not been
directly measured.  A second possibility is that \NeI\ is produced in
the heliosheath, perhaps from charge exchange between \NeII\ and \HeI,
since \HeII\ is abundant and the ionization potentials of \NeI\ and
\HeI\ are within 3 eV of each other.  We are unaware of any data on
the cross section for this reaction so we cannot evaluate this
possibility though it certainly warrants further study.

Models 2 and 8 show that Ar is $\sim$20\% neutral at the solar
location.  The radiation field is poorly constrained at the FIP of
\ArI\ (15.8 eV, $\lambda \sim$790 \AA) and highly sensitive to the cloud
\HI\ distribution.  The models and data are barely consistent if
filtration is modest.

The predicted depletions for Fe, Mg, and Si for Models 2 and 8 are
--1.18, --0.87, and --0.75, compared to H01 solar values.  These
depletions are comparable to values between warm and cool disk cloud
depletions \citep[e.g.,][]{Welty23:1999}.  Generally most depletions
are evaluated with respect to \HI\ rather than \HI+\HII as done here.
Had only \HI\ been considered, these ions would be $\sim$0.2 dex less
depleted, placing them closer to the warm disk gas
\citep[][]{Welty23:1999}.  For an ISM abundance pattern that is 70\%
of solar (using H01), Models 2 and 8 predict Fe, Mg, and Si depletions
of --1.02, --0.70, and --0.60 respectively.  For subsolar abundances,
C, N, O, and S are essentially undepleted, indicating extensive
destruction of small dust grains and grain mantles in the local ISM.

\subsection{Gas-to-Dust Mass Ratio \label{sec:rgd}}  

Paper I showed that \Rgd\ determined from \emph{in~situ} observations
of interstellar dust within the solar system is lower by factors of
1.5--6 when compared to \Rgd\ determined from missing mass arguments
applied to the nearby ISM.  Some of the ISM data used in that paper
have been updated (e.g. newer results for \epsCMa), and we revisit the
conclusions of Paper I using the results of Paper II.  The \Rgd\
predicted by applying the missing-mass arguments to the results of
Models 2 and 8 is now redetermined by assuming both solar (H01, GS98)
and 70\% of solar abundances for the ISM abundance pattern, and
including only elements listed in Tables \ref{table5} and
\ref{table6}.
H01 solar abundances yields \Rgd=178 and
183 for Models 2 and 8, with the increase compared to Paper I result primarily from a $\sim$20\%
lowering of the solar O abundance.  (For comparison, \Rgd\ for Models
11, 17, and 19 gives 163, 368, and 361 using GS98 abundances.) For an
ISM abundance pattern that is 70\% of the H01 value, Models 2 and 8
yield \Rgd=611 and 657.  If all trace elements heavier than He are
incorporated into the dust grains, \Rgd$ \sim66$.  Uncertainties of
$<$13\% follow from the neglect of additional elements in grain mass
calculations.  The missing mass arguments give a sightline average
value for \Rgd.

ISDGs from the LIC have been directly observed with detectors on board the
Ulysses and Galileo satellites, where retrograde orbits characterize the
interstellar versus interplanetary populations \citep{Baguhletal:1996}.  The
total dust grain mass measured by Ulysses/Galileo is $6.2\times10^{-27}$ g \cc, extending
over the mass range 3$\times 10^{-15} - 10^{-9}$ g \citep[Paper I,][]
{Landgrafetal:2000}.  This density can be compared to the interstellar
gas density to obtain a second estimate of \Rgd\ for the LIC.  Models
2 and 8 yield \Rgd$_\mathrm{LIC}=115^{+16}_{-14}$.  Smaller charged
grains ($<10^{-13}$ g), which in principle sample the MRN distribution
\citep{Mathis:1977}, are prevented from reaching the inner solar
system ($\sim$5 AU) by Lorentz-force coupling to the solar wind as
well as exclusion at the heliopause via coupling to ionized LIC gas.
Thus estimates of \Rgd$_{\rm LIC}$ from the Ulysses/Galileo data are
an upper limit to the true LIC value.

Both the B-star abundances and solar abundances are inconsistent with
the direct dust measurements of \Rgd$_\mathrm{LIC}$ by
Ulysses/Galileo, in agreement with the conclusions of Paper I.  This
difference suggests that gas and dust are not fully mixed over
sub-parsec scale lengths.  \citet{GruenLandgraf:2000} suggest that the
small grain population is enriched by the destruction of larger grains
which originate in a population that is not dynamically coupled to the
cloud.  These grains would not contribute to visual reddening, and are
therefore not included in the relatively constant gas-dust ratio found
from comparisons of \ebv\ and \HI+\HH.  The Ulysses/Galileo grains
share the LIC velocity, so the large uncoupled grain destruction that
may be enriching the total dust mass at the solar location happened
long ago enough for the resulting grain fragments to recouple to the
gas.

In the Appendix we calculate \Rgd\ values for several clouds to nearby
stars ($<500$ pc) and show that \Rgd\ is proportional to the
percentage of the dust mass which is carried by Fe (\PFe).  This
correlation, shown in Fig. \ref{fig2}, indicates that grain
destruction in the ISM increases \Rgd\ by
removing C, N, O, Si, and S from the grain and leaving behind an
Fe-rich resiliant grain core, as found previously (e.g. Frisch et
al. 1999).  In \S \ref{app:rgd} we find that \Rgd\ for the LIC is a
factor of 2--3 larger than \Rgd\ for the blue-shifted cloud,
indicating grain properties are inhomogeneous over parsec-sized
scales.  This grain destruction scenario has left behind Fe-rich grain
cores (younger than the $\sim10^{7-8}$ years required to replenish
grain masses either through depletion onto grain surfaces or
coagulation) in the LIC towards $\epsilon$ CMa.  Following the
arguments of \citet{GruenLandgraf:2000}, the larger interstellar grain
population observed by Ulysses and Galileo, in contrast, must consist
of grain fragments small enough to couple recently to the local
magnetic field, even though the parent unshattered grains and gas were
not kinematically coupled.  These large uncoupled grains are not
counted when missing mass arguments are used to calculate \Rgd.  Most
of the mass of the interstellar dust grains detected within the solar
system is carried by grains with masses $> 10^{-13}$ g (or radius
$>0.2$ $\mu$m), which have a gyro-radius in a weak magnetic field
($\sim$3 $\mu$G) of $\sim$0.3 pc \citep{GruenLandgraf:2000}.  This
distance is less than the distance to the LIC edge in the upstream
direction.  Theoretical studies predict that, because of the
differences in the creation and destruction timescales for
interstellar dust grains, \Rgd\ over the small scales (such as sampled
by the \emph{in~situ} data) differs from average values over longer
interstellar sightlines.

Estimates of depletions in the nearby ISM using data towards 21 stars
averaged together have yielded the result that the nearest ISM has
\Rgd$ = 73-151$, \citep{Kimuraetal:2002}, depending on the assumption
of \NHII/\NHI.  They also used somewhat different assumptions for the
reference ISM standard, including solar abundances based on (Pr02) data
which we have found inconsistent with our best models.  Our study here
(and \S \ref{app:rgd}) calculates \Rgd\ for each individual sightline,
with data on \CII, \NI, \OI, \MgII, \SiII, and \FeII, and in most
cases \SII, so these results can not be directly compared to the
Kimura et al. results.

\subsection{Interstellar Magnetic Field and Cloudlet Mass \label{sec:B}}  

In the conductive interface model \citep[][and Paper II]{Slavin:1989},
magnetic pressure in the interface (balanced by the pressure of the hot
plasma) is required to maintain interface EUV emission at levels which
reproduce the observed ionization.  Observations of the interstellar magnetic
field directly outside of the heliosphere have been elusive.  However, the
very weak polarization of the light from nearby stars suggest a weak nearby
magnetic field aligned in the plane of the Galaxy and directed towards a
Galactic longitude of $l \sim$90\deeg\ \citep{Tinbergen:1982}.  The ordered
component of the interstellar magnetic field traced by pulsar dispersion
measures suggest this field is weak \citep[$<3 \mu$G,][]{Frisch:1990}.  
Figure \ref{fig3} shows the electric vector polarization direction for several nearby stars
sampling the Tinbergen polarization patch.
The region of maximum polarization closely follows the ecliptic plane,
but also coincides with the LSR upstream direction of the CLIC.  The
classical interstellar dust grains which cause the polarization are
charged, and pile up in the heliosheath regions as they are deflected
around the heliosphere (Paper I), so they were not sampled by the
Ulysses/Galileo satellites.  The pileup of ``classical'' interstellar
grains in the heliosheath regions might possibly contribute to the
weak polarization observed by Tinbergen.

A magnetic field in the Galactic plane would be tilted with respect to
the ecliptic by $\sim$60\deeg.  The north ecliptic pole is directed
towards $l$=96\deeg, $b$=30\deeg, giving an ecliptic plane tilted by
$\sim$60\deeg\ with respect to the Galactic plane.  Observations by
Voyager 1 and 2 of a dozen $\sim$3 kHz emission events in the outer
heliosphere also suggest that the interstellar magnetic field
direction is parallel to the Galactic plane
\citep[][]{KurthGurnett:2003}.

The relatively good correlation between interstellar gas (\HI\ + \HH)
and color excess (\ebv) indicates that gas and dust are coupled on
scales of $\sim$100 pc or greater \citep{BohlinSavageDrake:1978}.  The
CLIC gas towards $\epsilon$ CMa, summing the two cloudlets together,
is very low mass.  Models 2 and 18 can be used to estimate the mass of these
cloudlets, treating them as a single spherical cloudlet.  This
cloudlet extends $<10,000$ AU towards $\alpha$ Cen ($\sim30$\deeg\
from the LIC LSR upstream direction), and $\sim1$ pc towards
$\epsilon$ CMa.  Assuming a nominal cloud diameter of 1 pc, as consistent 
with data, the
cloudlet mass is then $\sim$0.003 M$_\sun$.

\subsection{Origins of LIC \label{origins}}

The abundance pattern of refractory elements in the CLIC has been used as an
indicator of cloud origin \citep{Frisch:1981}.  The abundances of Si, Mg and
Fe in the CLIC gas are consistent with grain processing through a shock of
velocity $\sim100$ \kms\ (Paper I).

\citet{Frisch:1979,Frisch:1981} proposed that the nearest ISM
originated from the asymmetric expansion of a supershell, whose
kinematics were dominated by material flowing from higher density star
formation regions into the lower density regions of the Local Bubble,
and the refractory elements in the CLIC are consistent with processing
through a $\sim100$ \kms\ shock front (Paper I).  An alternative
scenario is that the CLIC originated as an outflow of material
evaporated from dense clouds in the Scorpius star formation region
\citep{Frisch:1995}, although this does not explain enhanced
refractory abundances.  An alternative scenario is that the nearby ISM
results from Rayleigh-Taylor unstable material ejected from the
``interface'' between the Local Bubble and Loop I where grains would
also have been shocked \citep{Breitschwerdtetal:2000}.  The bulk ISM
flow past the Sun has velocity in the Local Standard of Rest of
$-17\pm5$ \kms, from the upstream direction $l \sim2.3$\deeg, $b
\sim-5.2$\deeg \citep{Frischetal:2002}.  For all of these scenarios,
the star formation epochs in the Sco-Cen Association are younger than
the timescales for replenishing dust grains, so the grain destruction
processes have not yet been balanced by grain growth locally.

\section{Conclusions}

The primary conclusions of this paper are as follows:
\begin{enumerate}

\item By using line of sight data towards nearby stars
\citep[including \emph{FUSE} and STIS
data,][]{Moosetal:2002,AndreHowketal:2003} combined with observations
of the ISM interaction products inside of the solar system to choose
among radiative transfer models, we are able to select two models
which yield very good fits to available data.  These models make a
range of predictions, including the chemical composition of the ISM
near the Sun, the filtration factor for \HI\ entering the solar
system, and the physical properties of the interstellar cloud
surrounding the solar system.  The most definitive discriminants among
models turned out to be the global O/H ratio ($\sim$400 PPM),
\MgII/\MgI\ towards $\epsilon$ CMa, and the \HI/\HeI\ ratios inside
and outside of the solar system.  Both \HI\ and \HII\ need to be
included when evaluating abundances of ions found in warm diffuse
clouds.

\item We find that the two best models (Models 2 and 8 of Paper II)
indicate that the chemical composition of the nearest ISM is likely to
be subsolar, or $\sim70$\% of the H01 solar values.  This conclusion
rests primarily on the inferred S abundance under the assumption
that S is undepleted, and the assumed O abundance which yields good
matches between ISM data inside and outside of the heliosphere.

\item For these same two models, the filtration factor for \HI\ in the
heliosheath regions is $\sim0.46$.  Filtration factors for N
($\sim0.85$) and O ($\sim0.6$) are also predicted (\S
\ref{sec:insitu}).

\item We show that these models give gas-to-dust mass ratios
calculated for the nearest ISM towards $\epsilon$ CMa from the missing
mass method of \Rgd $ = 178-183$, provided that the chemical
composition of the nearest ISM is solar.  
We note that for this model to be viable S must be incorporated into dust.
This \Rgd\ value differs by nearly
a factor of two from the \Rgd\ calculated directly from Ulysses and
Galileo observations of interstellar dust grains in the solar system
(\Rgd$\simeq 115$, consistent with Frisch et al. 1999).  This result
is consistent with predictions that the differences in the creation
and destruction timescales for interstellar dust grains will cause
\Rgd\ to vary over sub-parsec length scales
\citep{Dwek:1998,Hirashita:1999}, and the fact that the ISM near the
Sun is part of a dynamically active cluster of cloudlets flowing away
from the Sco-Cen Association \citep{Frischetal:2002}.  

\item We show that the assumption of solar abundances also yields a grain
composition whereby the percentage of the dust mass that is carried by iron is
directly correlated to \Rgd\ (\S \ref{app:rgd}).  The implication is
that the Fe forms a robust core that is not destroyed during grain processing
in the ISM.  Since dust mass is proportional to the ISM metallicity
\cite{Dwek:1998}, this apparent correlation deserves further investigation.

\item For an ISM reference standard that is 70\% of H01 solar values, as is
indicated by our Models 2 and 8, the \Rgd\
ratio is raised by a factor of three to \Rgd=611--657.  In this case we infer a
dust composition of primarily Fe, Mg and Si.  The correlation between \Rgd\
and the Fe fraction of the dust mass is preserved for this case, although
\Rgd\ values are increased.

\item Comparisons between \insitu\ dust data and these results suggest
nearby interstellar gas and dust haven't been fully coupled over the
lifetime of the cloud.  If either gas-dust coupling breaks down over
the cloud lifetime, or if refractory elements are also present in
subsolar abundances, then applying missing mass arguments to determine
dust grain mineralogy will not work.

\item The neon abundance remains a problem for these models, which may
indicate either incorrect solar abundances, incorrect ionization
correction (due to a poorly understood EUV radiation field), or
possibly unmodeled charge exchange between interstellar \HeI\ and
\NeII\ in the heliosheath regions.

\end{enumerate}

These results are encouraging since they show that radiative transfer
models based on accurate interstellar radiation field data, and
combined with precise ISM measurements towards nearby stars and inside
the heliosphere, offer the possibility of understanding the detailed
physics of the ISM and the interaction between the heliosphere and the
ISM.  However, the conclusion that Models 2 and 8 are the best models
is sensitive to the data uncertainties.  This paper as originally
submitted used data from \citet[][private
communication]{GloecklerGeiss:2001} and \citet[][private
communication]{Witte:1996}.  During the extended refereeing process
newer data became available and are incorporated into this paper.  The
differences between the H and He densities in the earlier data and the
values used here are $<$10\%, yet this difference is large enough to
replace Model 18 (as originally concluded) with Model 8 (as found
here) as one of the two best models.

\acknowledgements
The authors would like to gratefully acknowledge research support from NASA
grants, including NAG5-6405, and NAG5-11005, and NAG5-8163.  We thank Alan
Cummings, Ed Stone, George Gloeckler, Hans Mueller, Gary Zank, Dan Welty, and
Don York for many helpful discussions.

\section{Appendix}

\subsection{$\epsilon$ CMa Data \label{app:cma} }

The column densities for the d$<$3 pc LIC and blue-shifted clouds 
towards $\epsilon$ CMa were summed together in Paper II, since both
clouds appear to contribute to the attenuation of the interstellar
radiation field.
The resulting column densities (from Paper I) are:
\NCII= $2.1 - 3.4\times10^{14}$ \cmtwo,
\NCIIstar= $1.5 \pm 0.31 \times10^{12}$ \cmtwo, 
\NNI= $2.68 \pm 0.1 \times10^{13}$ \cmtwo,
\NOI= $2.6^{+0.8}_{-0.5} \times10^{14}$ \cmtwo,
\NMgII=$4.15 \pm 0.11 \times10^{12}$ \cmtwo,
\NMgI=$1.2 \pm 0.3 \times10^{10}$ \cmtwo,
\NSiII= $6.37\pm 0.3 \times10^{12}$ \cmtwo,
\NSII= $1.35 \pm 0.36 \times10^{13}$ \cmtwo,
and \NFeII= $1.87\pm 0.1 \times10^{12}$ \cmtwo.

\subsection{Solar Abundances \label{sec:Asolar}}

Solar abundances have been established by GS98, using a combination of
photospheric, corona, and meteoritic abundances.  (Abundances are
summarized in Table \ref{table1}.)  Photospheric data provide C
($331\pm49$ PPM), N ($83\pm12$ PPM), and O ($676\pm100$ PPM)
abundances.  Less precise coronal data give Ne ($120\pm18$ PPM) and Ar
($2.51\pm0.37$ PPM).  H01 updated the photospheric abundances,
obtaining new C ($391\pm110$ PPM), N ($85\pm25$ PPM), O ($545\pm107$
PPM), and Neon ($100\pm17$ PPM) abundances.  Pr02 considered solar
granulation and found solar C, N and O abundances lower than those of
GS98 or H01 for C ($245^{+24}_{-21}$ PPM), O ($490^{+136}_{-97}$ PPM),
and for N (68 PPM).

C, N, O.  \citet{Holmgrenetal:1990} determined B-star Ar abundances of
$3.16^{+0.39}_{-0.34}$, which is larger than, but within the
uncertainties of, the solar abundance (Table \ref{table1}).
In this paper we adopt H01 and GS98 values for solar abundances with the H01
values given priority.

\subsection{Model Predictions \label{app:predictions}}

The predictions of the 25 models in Paper II are shown, by model number, in Figs. 
\ref{fig4}--\ref{fig6}.
Fig \ref{fig4} shows the ionization diagnostics, \MgII/\MgI\ and \CII/\CIIstar, for each model, and the value towards $\epsilon$ CMa.
Models 2, 8, and 18 provide the
best fit to both the Mg and C ionization diagnostics.  
The predicted abundances (in PPM) for S, C, N, and O are shown in
Fig. \ref{fig5}.  The interstellar O abundance is plotted as dashed lines.  

Fig. \ref{fig6} shows the model predictions for He, Ar, Ne,
O, N, and T at the entry point to the heliosphere.
The uncertainties on these observation values are the \insitu\ data
are also shown.  The Ar values differ from values in Paper II because 
an abundance of 3.16 PPM is used.

\subsection{Gas-to-Dust Mass Ratio towards Nearby Stars \label{app:rgd}}
 
\begin{figure}[ht]
\vspace*{5.8in}
\includegraphics{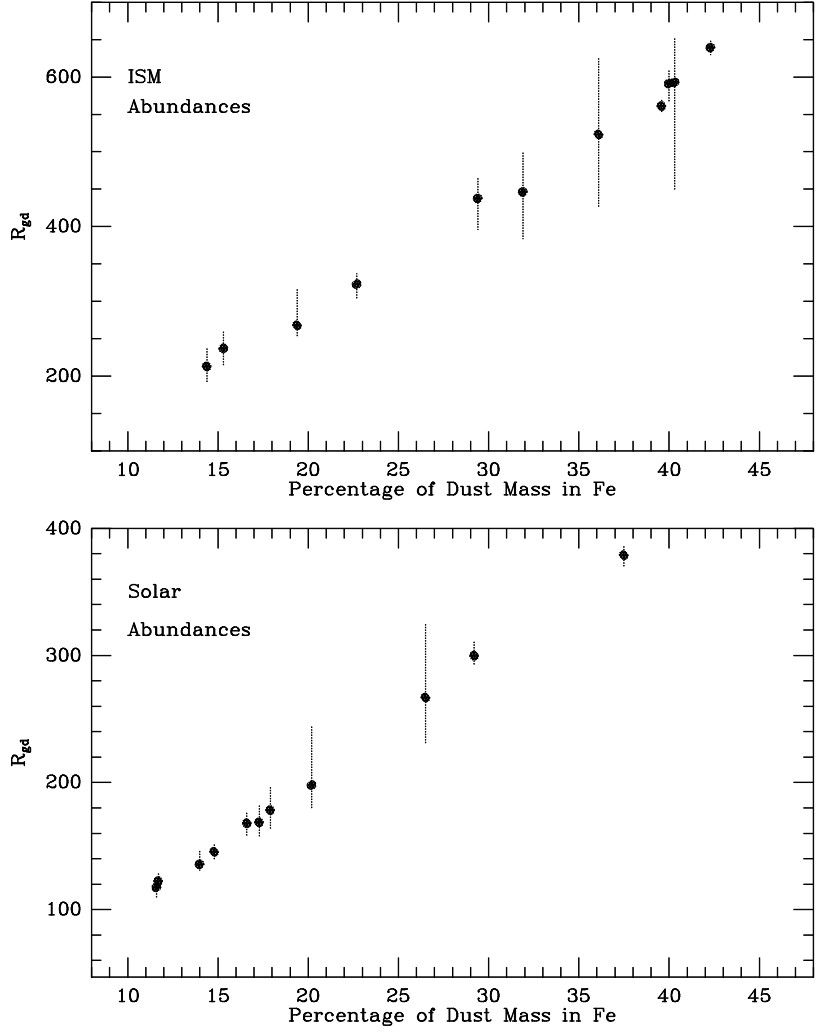}
\caption{ Plot of the gas-to-dust mass ratio (\Rgd) versus the
percentage of dust grain mass carried by iron, towards a set of clouds
seen towards relatively nearby stars.  The uncertainties on \Rgd\
values are shown.  Data for clouds towards $\zeta$ Oph (both warm and
cold clouds), $\eta$ UMa, Capella, $\lambda$ Sco, and 23 Ori (both
`WL' and `SL' components) are plotted.  The LIC and blue-shifted
components towards Sirius and $\epsilon$ CMa are also plotted.  The
bottom plot shows \Rgd\ values calculated for a reference standard in
the ISM that is equal to solar abundances (using H01 and GS98
abundances).  The top plot shows \Rgd\ values calculated for a
reference standard that is 70\% of solar values in the ISM.  The
Appendix (\S \ref{app:rgd}) lists the data sources for this plot.  }
\label{fig2}
\end{figure}

\begin{figure}[ht]
\vspace{4.in}
\includegraphics{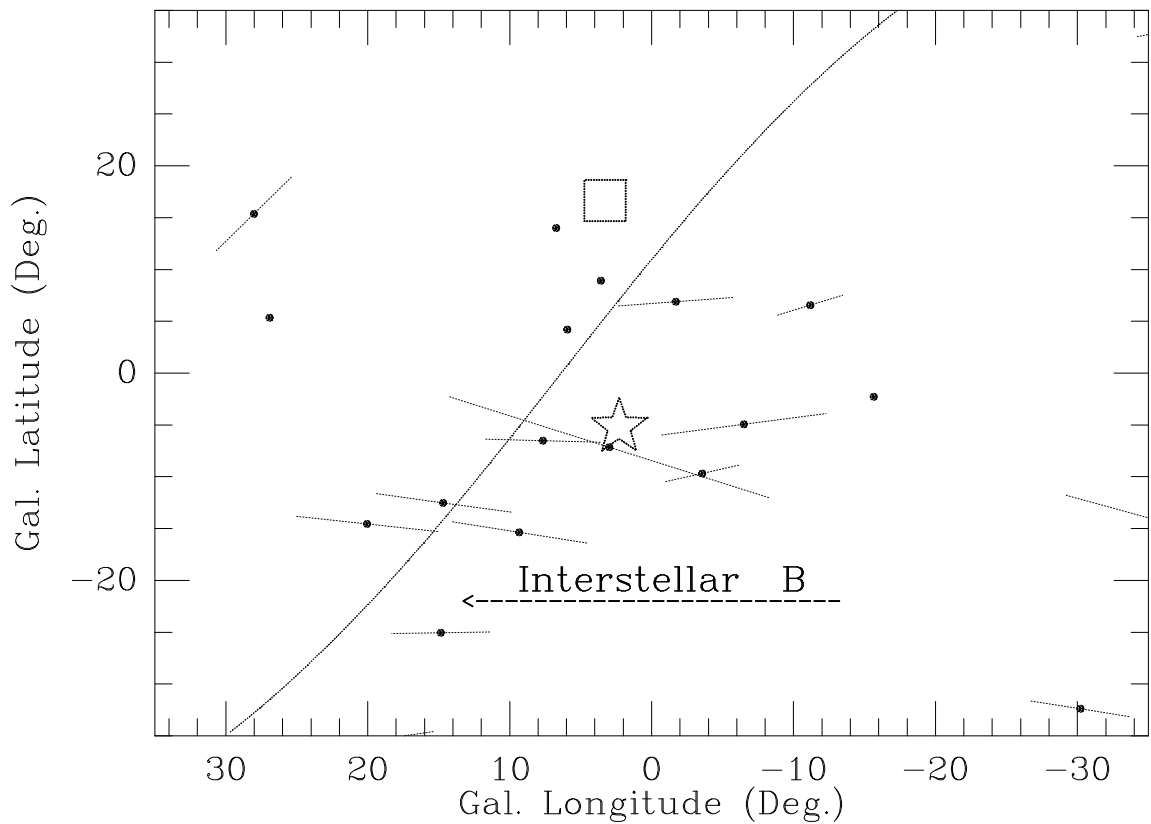}
\caption{Plot of an indicator of the nearby interstellar magnetic field, in
galactic coordinates.
The bars show the direction of the electric vector polarization, which
is parallel to the interstellar magnetic field direction, for several
nearby stars (Tinbergen 1982).  The arrow shows the likely direction
of the interstellar magnetic field near the Sun based on this data.
The curved line shows the ecliptic plane.  The region of maximum
polarization follows the ecliptic plane.  The classical interstellar
dust grains which polarize optical radiation pile up in the
heliosheath regions as they are deflected around the heliosphere
(Frisch et al. 1999).  
The box shows the heliosphere nose direction in heliocentric coordinates,
and the star shows the CLIC bulk flow upstream direction in the the
local standard of rest (see text).  }
\label{fig3}
\end{figure}

\begin{figure}[ht]
\plotone{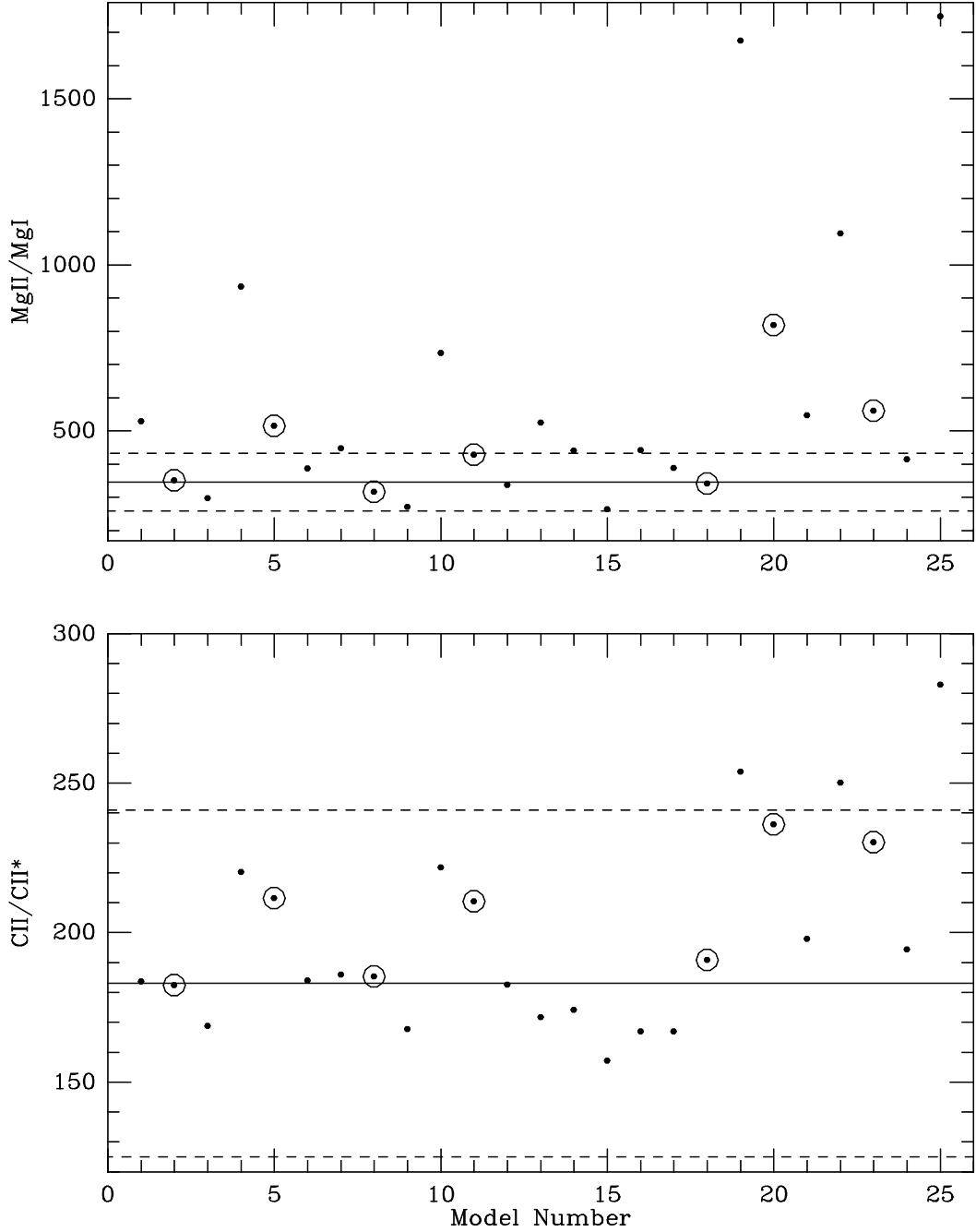}
\vspace*{-0.8in}
\caption{
Predicted \NMgII/\NMgI\ (top) and \NCII/\NCIIstar\ (bottom) plotted against
model number.  The observed ratios towards $\epsilon$ CMa are plotted as lines
(with the uncertainties plotted as dashed lines).  
The models consistent with O/H$\sim$400 PPM are circled.
}
\label{fig4}
\end{figure}

\begin{figure}[ht]
\vspace*{-3.0in}
\plotone{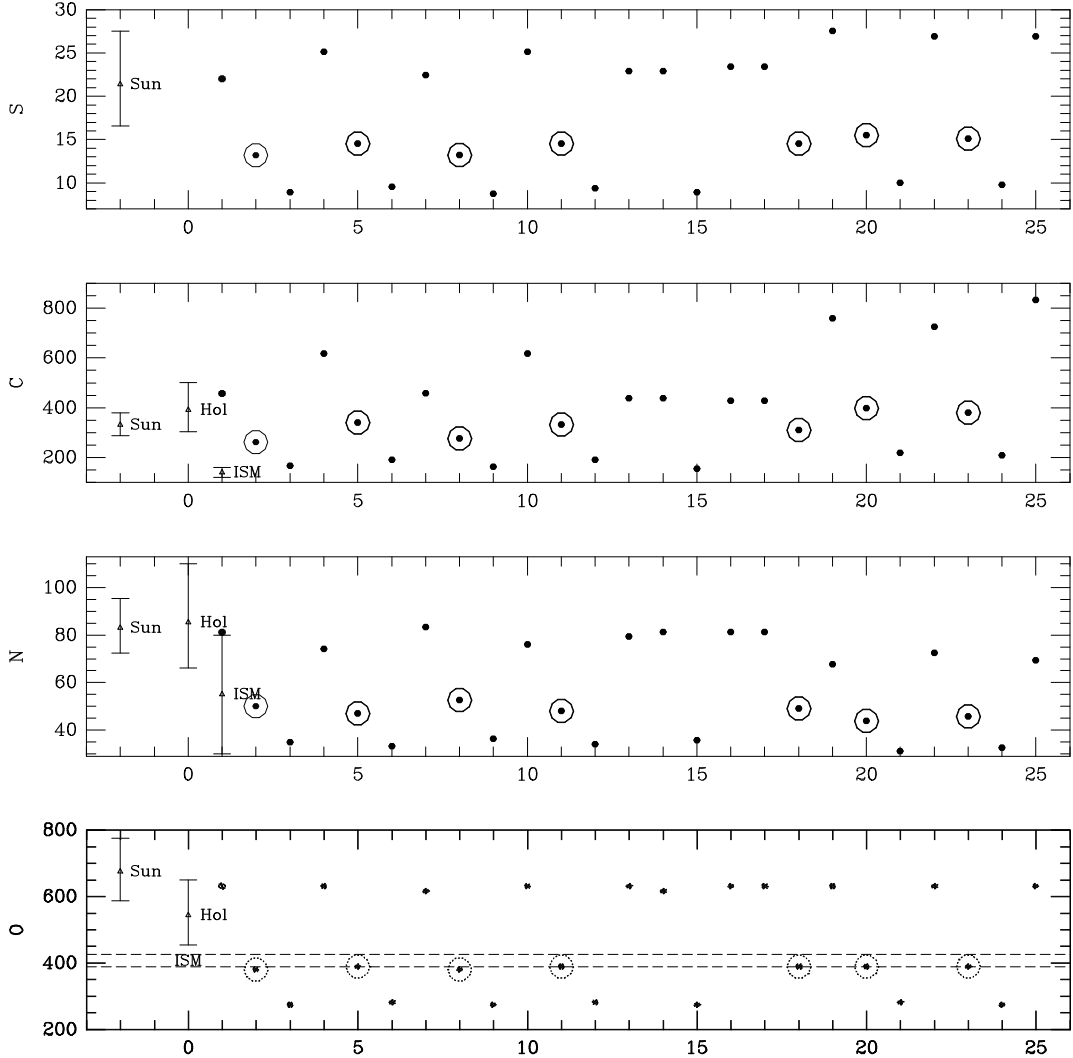}
\vspace*{-0.8in}
\caption{
Plot of S, C, N, and O (top to bottom) abundances predicted for the 25 Models,
shown in PPM versus model number.  Comparison abundances are shown for solar
abundances (Sun, GS98), H01 abundances and the ISM values
\citep{AndreHowketal:2003,SofiaMeyer:2001}.  
The models consistent with O/H$\sim$400 PPM are circled.
}
\label{fig5}
\end{figure}

\begin{figure}[ht]
\vspace*{-0.4in}
\plotone{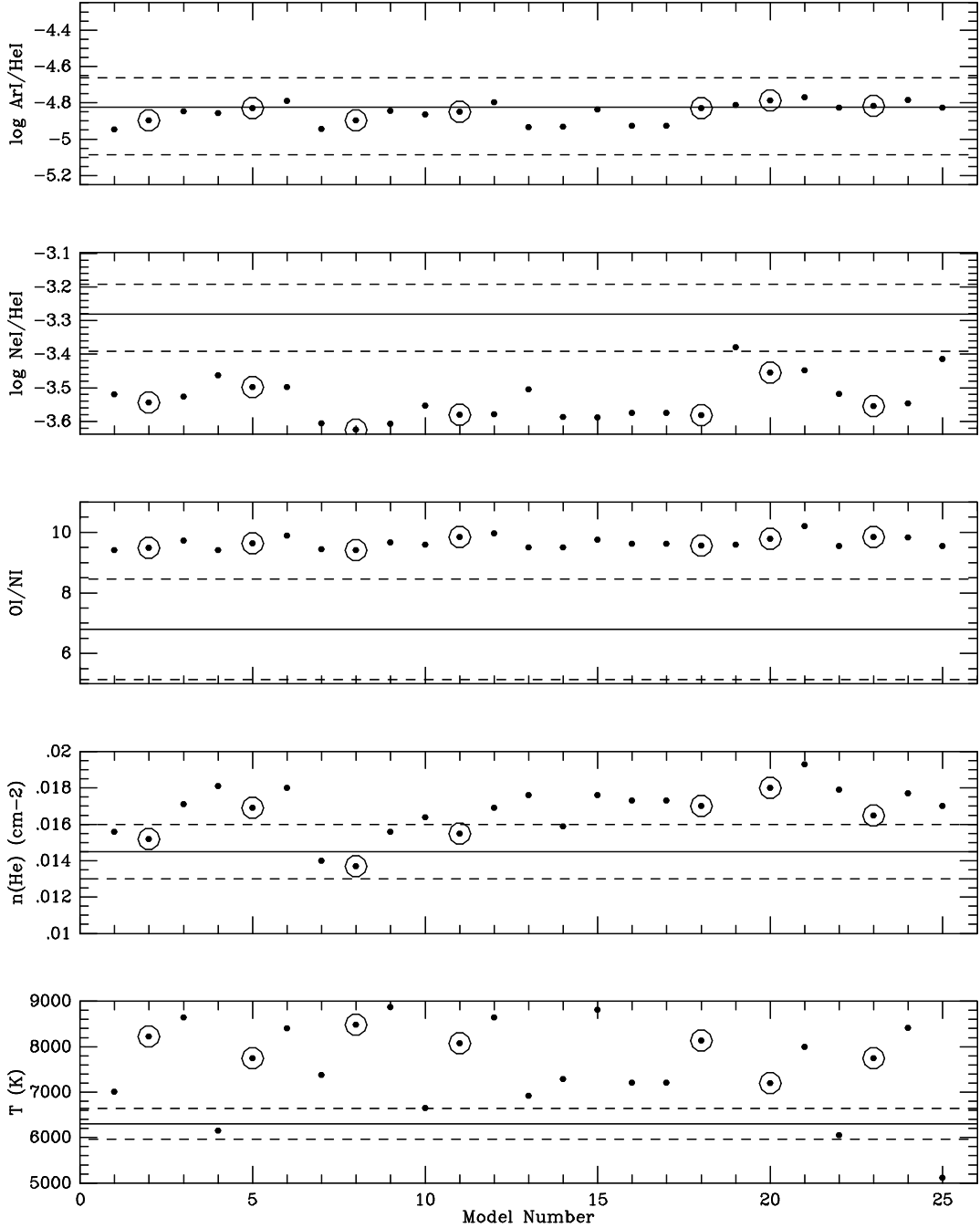}
\vspace*{-1.0in}
\caption{ Predicted quantities for the solar location are plotted
against model number.  The comparison data, which includes pickup ion,
direct observations of \HeI, and anomalous cosmic ray data (Table
\ref{table3}), are plotted as lines (with uncertainties shown as
dashed lines.)  Top to bottom: log(\ArI/\HeI), log(\NeI/\HeI), \OI/\NI,
\nHeI, and temperature.  The models consistent with
O/H$\sim$400 PPM are circled.  }
\label{fig6}
\end{figure}

For Fig. \ref{fig2}, \Rgd\ was calculated
as described in \S \ref{sec:rgd} and \citet{Frischetal:1999}.
References for the data used to construct Fig.  \ref{fig2} are as
follows: warm and cold clouds towards $\zeta$ Oph
\citep{SavageSembach:1996,SavageCardelliSofia:1992,CardelliSavageEbbets:1991,Federmanetal:1993,Morton:1975},
$\eta$ UMa (Frisch et al. unpublished), $\lambda$ Sco
\citep{York:1983}, and 23 Ori \citep[both `WL' and `SL'
components][]{Welty23:1999}, and the LIC and blue-shifted clouds
towards $\alpha$ CMa \citep{Hebrardetal:1999}, and $\epsilon$ CMa
\citep{GryJenkins:2001}.  Capella results are also included
\citep{WoodRedfieldetal:2002}.  For all of these clouds, data on \CII,
\NI, \OI, \MgII, \SiII, and \FeII\ are available, and in most cases
also \SII\ data.  In some cases, \HI\ and \HII\ were estimated from
ionization indicators such as \NI\ and \NII, while towards the two CMa
stars the two individual clouds were assumed to have
ionization levels predicted by Models 2 and 8.
The bottom plot shows \Rgd\ values calculated for an assumed
chemical composition standard that is equal to the solar abundance
pattern (using H01, when available, and GS98 abundances).  The top
plot shows \Rgd\ values calculated using a reference standard in the
ISM that is 70\% of the solar values.  The correlation between \Rgd\
and the percentage of grain mass carried by Fe appears to indicate
that grain destruction increases \Rgd\ and leaves behind an iron-rich
grain core.  The clouds with \PFe $>$ 25 include the LIC clouds
towards Sirius and $\epsilon$ CMa, $\lambda$ Sco, $\zeta$ Oph warm
cloud, and the two clouds towards 23 Ori.  All of these clouds have a
known origin in star-formation regions, and except for the 23 Ori SLV
cloud, relatively large velocities ($|V| >$15 \kms) in the LSR.  The
blue-shifted clouds towards Sirius and $\epsilon$ CMa have \PFe $<$25,
and show \Rgd\ values that are factors of 2--3 less than for the LIC.
Evidently grain destruction is minimized in blue-shifted cloud,
suggesting different origins for the LIC and blue-shifted clouds
\citep[e.g.][]{Frisch:1995}.  Note that He is included in gas-mass
estimates, and \OI\ and \NI\ abundances are referenced to \HI\ alone
for calculating the points shown in Fig. \ref{fig2}.

\subsection{Predicted Ionization Levels for Models 2, 8, and 18 }

The ionization levels predicted for H, He, C, N, O, Ne, Na, Mg, Al,
Si, P, S, Ar,Ca, and Fe by Models 2, 8, and 18 are listed in Tables
\ref{table5}, \ref{table6}, and \ref{table7}.

\clearpage
\begin{deluxetable}{l lll ccccccc l cccc}
\tablecolumns{16}
\tabletypesize{\footnotesize}
\rotate
\tablewidth{0pc}
\tablecaption{Abundances:  Solar, ISM, and Models\tablenotemark{a}\label{table1}}
\tablehead{
\colhead{Element} & \multicolumn{3}{c}{Solar Reference Abundance\tablenotemark{b}} &\multicolumn{7}{c}{(\Xtot/\Htot\ from SF02\tablenotemark{c}$~$ Models)} & \multicolumn{5}{c}{\Xn/\HI\ in Gas Phase ISM\tablenotemark{d}}  \\

\colhead{} & \colhead{GS98} & \colhead{H01} & \colhead{Pr02} & \colhead{2} & \colhead{5} & \colhead{8} & \colhead{11} & \colhead{18} & \colhead{20} & \colhead{23} &
\colhead{Ion} & \colhead{Ref.} & \colhead{$\epsilon$ CMa\tablenotemark{f}} & \colhead{Capella\tablenotemark{g}} & \colhead{HZ43\tablenotemark{h}} \\ 
\cline{12-16} \\
} 
\startdata
C  &     331$^{+49}_{-43}$ & 391$^{+110}_{-86}$ & 245$^{+24}_{-21}$& 263 & 339 & 275  &331 & 309 & 398 & 380 & \CII &
141$^{+21}_{-18}$\tablenotemark{e}& 423$\pm$102 & 363:: & 794::  \\
N  &     83$^{+12}_{-11}$ &   85$^{+25}_{-19}$& $\sim$68 & 50.1  & 46.8 & 52.5 & 47.9 & 49.0 & 43.7 & 45.7 & \NI & 75$\pm$4\tablenotemark{e}& 41$\pm$2 & 42$\pm$2 & 38$\pm$8 \\
O  &     676$^{+100}_{-87}$ & 545$^{+107}_{-90}$ & 490$^{+60}_{-53}$ &380 & 389 & 380 & 389 & 389 & 389 & 389 & \OI & 375$\pm$47\tablenotemark{j}& [400$\pm$126] & 490:: & 380$\pm$87  \\
Ne &     120$^{+18}_{-16}$ &   100$^{+17}_{-15}$ & \nodata & [123] &  \nodata &  \nodata & \nodata &
 \nodata & \nodata & \nodata & \nodata &  \nodata& \nodata &  \nodata & \nodata \\
Mg &38$^{+5}_{-4}$& 35$^{+5}_{-4}$  & \nodata &4.68 &5.25 &4.90 &5.37 &5.37 &5.62 &5.75 & \MgII & \nodata & 6.4$\pm$0.2 & 3.7&  3.0 \\
Si &34$\pm$4& 34$\pm$4 & \nodata &6.03 &6.76& 6.03 & 6.61 &6.61 &7.08 & 6.92 & \SiII &  \nodata& 10$\pm$1 &5.34 & 8.3 \\
S  &     21$^{+6}_{-5}$  & 22$\pm$5  &  \nodata & 13.2 & 14.5 & 13.2   & 14.5& 14.5 & 15.5 & 15.1 & \SII & \nodata & 21$\pm$6  & $<$23 & \nodata   \\
Ar &     2.51$^{+0.37}_{-0.33}$ &  \nodata & (3.16$^{+0.39}_{-0.34}$)  &[2.82] & \nodata & \nodata & \nodata  & \nodata & \nodata & \nodata & \ArI & \nodata & \nodata & $<$18 & 1.1  \\
Fe &     31.6$^{+3.9}_{-3.4}$ &  28$^{+6}_{-5}$ & \nodata &1.86 & 2.04 & 1.86 & 2.00 & 2.04 & 2.14  & 2.09 & \FeII & &
 2.9$\pm$0.2 &  1.8 & 1.7 \\ 
\enddata
\tablenotetext{a} {All abundances are given in parts per million H atoms (PPM).}
\tablenotetext{b} { Solar abundances are from \citet[][GA98]{GrevesseSauval:1998}, \citet[][H01]{Holweger:2001}, and \citet[][Pr02]{PrietoAsplund:2002}.  The Ar abundance
enclosed in parentheses in the Pr02 column is from B-star abundances determined
by \citet{Holmgrenetal:1990}.}
\tablenotetext{c} {
The SF02 models (columns 5-11) present ISM abundances with respect to \HI+\HII\ for the $\epsilon$ CMa sightline.
Model abundances in brackets are assumed and invariant between the models.  
}
\tablenotetext{d} {
The ``Reference'' column refers to global ISM studies which include
dense cloud sightlines.
Interstellar values marked with ``::'' have uncertainties on the order of 100\%.}
\tablenotetext{e} { Reference ISM abundances from \citet[][C, N]{SofiaMeyer:2001}.}
\tablenotetext{f} {Based on \NHI=6.5 x 10$^{17}$ \cmtwo.}
\tablenotetext{g} {Data from \citet{WoodRedfieldetal:2002}.}
\tablenotetext{h} {Data from \citet{Kruketal:2002}, using curve of growth
values in Table 5.}
\tablenotetext{i} {Based on \NHI=6.37 x 10$^{17}$ \cmtwo, from O/H=408.}
\tablenotetext{j} {We combine O/H=408$\pm$14 PPM \citep{AndreHowketal:2003}, and O/H=343$\pm$15 
from GHRS data \citep{MeyerJuraCardelli:1998} to obtain the listed value 375$\pm$47 (\S \ref{sec:ism}).}

\end{deluxetable}

\begin{deluxetable}{ll ccccc cc}
\tablecolumns{9}
\tabletypesize{\footnotesize}
\tablewidth{0pc}
\tablecaption{Ionization Diagnostics\tablenotemark{a}\label{table2}}
\tablehead{
\colhead{Ratio} & \colhead{Value\tablenotemark{b}} &\multicolumn{7}{c}{SF02 Models\tablenotemark{c}}  \\
\colhead{} & \colhead{} & \colhead{2} & \colhead{5} & \colhead{8} & \colhead{11} & \colhead{18} & \colhead{20} & \colhead{23}  \\
\colhead{} & \colhead{} & \colhead{*} & \colhead{} & \colhead{*} & \colhead{} & \colhead{*} & \colhead{} & \colhead{}  \\
}
\startdata

\MgII/\MgI & 346$\pm$87 & 351.0  &  515.8 & 316.4 & 428.7 & 341.9 & 819.3 &  560.6 \\ 
\CII/\CIIstar & 183$\pm$58 &  182.4  &  211.5 & 185.4  & 210.5 & 190.9 & 236.2 & 230.3
 \\ 
\enddata
\tablenotetext{a} { Abundances given in parts per million H atoms (PPM).}
\tablenotetext{b} { For the combined LIC and blue-shifted cloud in the $\epsilon$ CMa sightline.}
\tablenotetext{c} { Best models based on $\epsilon$ CMa Mg and C data.}
\end{deluxetable}


\begin{deluxetable}{lllll lllll l}
\tablecolumns{11}
\rotate
\tablewidth{0pc}
\tablecaption{Interstellar Gas at Solar Location:  $In~situ$ Data versus Models \label{table3}}
\tablehead{
\colhead{Quantity} & \colhead{Results\tablenotemark{1}} & \colhead{Ref.} & \multicolumn{7}{c}{Model Predictions (no filtration)} & \colhead{Filtration\tablenotemark{2}} \\
\colhead{} & \colhead{} & \colhead{}& \colhead{2} & \colhead{5} & \colhead{8} & \colhead{11} & \colhead{18} & \colhead{20} & \colhead{23}  \\
\cline{4-10} \\
} 
\startdata

\HI\ (\cc)& 0.095$\pm$0.01  & 4 &  0.208 & 0.225 & 0.202 & 0.212 & 0.242 & 0.228 & 0.216 &0.40, (0.43)  \\
\HI\      &                 & 6 &0.213 & 0.213 & 0.236 & 0.218 & 0.228 & 0.203 & 0.209 & From text  \\
\HeI\ (\cc)& 0.0145$\pm$0.0015  & 3,4 &  0.015 & 0.017 & 0.014 & 0.016 & 0.017 & 0.018 & 0.017 & 0.99, 0.94  \\
\NI/\HeI\ & (5.38$\pm$1.17)E-4 & 4  
  & 5.63E-4 & 5.49E-4 & 6.12E-4 & 5.53E-4 & 5.92E-4 & 5.11E-4 & 5.28E-4 & 0.96, 0.76 \\
\OI/\HeI\ & (3.66$\pm$0.67)E-3 & 4,5 & 
  5.34E-3 & 5.29E-3 & 5.76E-3 & 5.45E-3 & 5.65E-3 & 5.00E-3 & 5.20E-3 & 1.17, 0.70 \\
\NeI/\HeI & (5.24$\pm$1.18)E-4 & 4 &
  2.85E-4 & 3.18E-4 & 2.37E-4 & 2.62E-4 & 2.62E-4 & 3.50E-4 & 2.79E-4  & 0.87, 0.88 \\
\ArI/\HeI & (1.50$\pm$0.67)E-5 & 5,4 & 
  1.07E-5 & 1.27E-5 & 1.08E-5 & 1.20E-5 & 1.25E-5 & 1.40E-5 & 1.30E-5 & 0.90, 0.64 \\
Temperature & 6,300$\pm$340 & 3  &  8,230 & 7,750 & 8,480 & 8,080 & 8,140 & 7,200 & 7,750 \\
  (K) &&&&&&&&&\\

\enddata
\tablenotetext{1} { ``E-n'' indicates multiplication by 10$^{-n}$.}
\tablenotetext{2} { Filtration factors are from \citet[][Model 1, upwind TS]{MuellerZank:2002,CummingsStone:2002}, and \citet[][\HI\ only]{Izmodenovetal:1999b}.  The filtration factors for ratios refer to the numerator
(the \HeI\ filtration is not repeated).  The \HI\ filtration factor in parentheses
is derived in the text from Model 18.} 
\tablenotetext{3} {The listed \nHeI\ value is the average of the Ulysses and pickup ion
(next note) measurements of \nHeI.  The final values from the 12-year Ulysses mission 
observations of interstellar \nHeI\ in the solar system give a velocity vector of
26.3$\pm$0.4 \kms\ towards downstream direction $\lambda$=74.7$\pm$0.5\deeg, 
$\beta$=--5.2$\pm$0.2\deeg, corresponding to an upstream direction of
\glong=3.3\deeg, \glat=+15.9\deeg\ \citep{Witteetal:2003}.}
\tablenotetext{4} {Pickup ion data are values at the termination shock from GG03.  }
\tablenotetext{5}{Anomalous cosmic ray data from \citet{CummingsStone:2002}, for
values at termination shock.}

\tablenotetext{6}{These values for \nHI\ are derived using the filtration
factor formula of \citet{GloecklerGeiss:2001}, with \nis\ values from 
the models.  See Section \ref{sec:filt}.}

\end{deluxetable}

\begin{deluxetable}{lccc}
\tablecolumns{4}
\tabletypesize{\normalsize}
\tablewidth{0pc}
\tablecaption{Results of Models\label{table4}}
\tablehead{
\colhead{Quantity} & \colhead{Model 2} & \colhead{Model 8}  &\colhead{Model 18}  \\
}
\startdata

\sidehead{$Assumed~Parameters:$}
$n_{\rm H}$ (\cc)& 0.273 & 0.273 & 0.300 \\
log $T_{\rm h}$ (K) & 6.0  & 6.1 & 6.1 \\
$B_{\rm o}$ ($\mu$G) & 5.0 & 5.0 & 3.0 \\
$N_{\rm HI}$ (10$^{17}$ \cmtwo) & 6.5 & 6.5 & 6.5 \\
FUV Field\tablenotemark{a} & MMP & MMP & GPW \\
\sidehead{$Predicted~Properties~for~ISM~at~Solar~Location:$}
\nHI\ (\cc) & 0.208 & 0.202 & 0.242 \\
\nHeI\ (\cc) & 0.016 & 0.014 &  0.017 \\
\ne\ (\cc) & 0.098 & 0.10 & 0.089 \\
$\chi$(\HI)\tablenotemark{b} & 0.287 & 0.30 & 0.234 \\
$\chi$(\HeI) & 0.471 & 0.51 &  0.448 \\
T (K) & 8,230 & 8,480 &  8,140 \\
\Rgd -- Ulysses/Galileo Data & 103 & 113 \\
\sidehead{$Predicted~Properties~for~ISM~in~ \epsilon~CMa~Sightline:$}
log \NHIHII\ (\cmtwo) & 18.03 & 18.02 &  17.98 \\ 
\NHI/\NHeI\ & 11.6 & 12.7 & 12.1 \\
\Rgd --Solar Abundances\tablenotemark{c} & 178 & 183 & 198 \\ 
\Rgd --ISM Abundances\tablenotemark{c} & 611 & 657 & 669 \\ 
 \\ 
\enddata
\tablenotetext{a} {MMP$\rightarrow$Mathis et al. 1983; \\
GPW$\rightarrow$Gondhalekar et al. 1980.}
\tablenotetext{b} {$\chi$(X) is the ionization fraction of element X.}
\tablenotetext{c} {The assumed solar abundances are from Holweger (2001, see text).
The ISM reference abundance is assumed at 70\% of the
Holweger (2001) solar values.}
\end{deluxetable}

\begin{deluxetable}{llllll}
\tablecolumns{6}
\tablewidth{0pc}
\tablecaption{Model 2$~~$Predictions for Ionization Fractions at the Sun\tablenotemark{a} \label{table5}}
\tablehead{
\colhead{Element} & \colhead{Abundance} & \multicolumn{4}{c}{Ionization
Fraction} \\
\colhead{} & \colhead{(ppm)} & \colhead{I} & \colhead{II} 
& \colhead{III} & \colhead{IV}}
\startdata
    H  & 1.00E+06 &    0.714  &   0.287 &     0.00 &     0.00  \\
    He & 1.00E+05 &    0.521  &   0.471 &  0.00838 &     0.00  \\
    C  &     263. & 0.000457  &   0.968 &   0.0319 &     0.00  \\
    N  &     50.1 &    0.586  &   0.414 & 0.000213 &     0.00  \\
    O  &     380. &    0.733  &   0.267 & 0.000124 &     0.00  \\
    Ne &     123. &    0.121  &   0.646 &    0.233 & 6.23E-06  \\
    Na &     2.04 &  0.00143  &   0.889 &    0.109 & 4.09E-06  \\
    Mg &     4.68 &  0.00214  &   0.825 &    0.173 &     0.00  \\
    Al &   0.0794 & 7.46E-05  &   0.974 &   0.0173 &  0.00894  \\
    Si &     6.03 & 4.64E-05  &   0.997 &  0.00336 & 3.19E-05  \\
    P  &    0.219 & 0.000173  &   0.976 &   0.0234 & 0.000102  \\
    S  &     13.2 & 8.51E-05  &   0.961 &   0.0384 & 3.01E-06  \\
    Ar &     2.82\tablenotemark{b} &    0.199  &   0.488 &    0.313 & 6.22E-06  \\
    Ca & 0.000407 & 2.92E-05  &  0.0178 &    0.982 & 0.000211  \\
    Fe &     1.86 & 0.000198  &   0.967 &   0.0332 & 9.59E-06  \\
\enddata
\tablenotetext{a}{Model 2 from \citet{SlavinFrisch:2002}.}
\tablenotetext{b}{This is the Ar abundance used in Paper II.  
In this paper we use Ar/H=3.16 PPM from \citet{Holmgrenetal:1990}.}
\end{deluxetable}

\begin{deluxetable}{llllll}
\tablecolumns{6}
\tablewidth{0pc}
\tablecaption{Model 8$~~$Predictions for Ionization Fractions at the Sun\tablenotemark{a} \label{table6}}
\tablehead{
\colhead{Element} & \colhead{Abundance} & \multicolumn{4}{c}{Ionization
Fraction} \\
\colhead{} & \colhead{(ppm)} & \colhead{I} & \colhead{II} 
& \colhead{III} & \colhead{IV}}
\startdata
    H  & 1.00E+06 &    0.701 &    0.299  &    0.00 &     0.00  \\
    He & 1.00E+05 &    0.475 &    0.511  &  0.0141 &     0.00  \\
    C  &     275. & 0.000449 &    0.961  &  0.0382 &     0.00  \\
    N  &     52.5 &    0.554 &    0.446  &0.000296 &     0.00  \\
    O  &     380. &    0.721 &    0.279  &0.000169 &     0.00  \\
    Ne &     123. &   0.0916 &    0.610  &   0.298 & 1.06E-05  \\
    Na &     2.04 &  0.00122 &    0.848  &   0.151 & 3.90E-06  \\
    Mg &     4.90 &  0.00232 &    0.791  &   0.206 &     0.00  \\
    Al &   0.0794 & 7.85E-05 &    0.973  &  0.0180 &  0.00900  \\
    Si &     6.03 & 4.80E-05 &    0.996  & 0.00401 & 3.39E-05  \\
    P  &    0.219 & 0.000177 &    0.972  &  0.0276 & 0.000124  \\
    S  &     13.2 & 8.71E-05 &    0.955  &  0.0454 & 4.46E-06  \\
    Ar &     2.82\tablenotemark{b} &    0.182 &    0.469  &   0.349 & 9.23E-06  \\
    Ca & 0.000407 & 3.33E-05 &   0.0181  &   0.982 & 0.000289  \\
    Fe &     1.86 & 0.000225 &    0.965  &  0.0353 & 1.27E-05  \\
\enddata
\tablenotetext{a}{Model 18 from \citet{SlavinFrisch:2002}.}
\tablenotetext{b}{This is the Ar abundance used in Paper II.  
In this paper we use Ar/H=3.16 PPM from \citet{Holmgrenetal:1990}.}
\end{deluxetable}

\begin{deluxetable}{llllll}
\tablecolumns{6}
\tablewidth{0pc}
\tablecaption{Model 18$~~$Predictions for Ionization Fractions at the Sun\tablenotemark{a} \label{table7}}
\tablehead{
\colhead{Element} & \colhead{Abundance} & \multicolumn{4}{c}{Ionization
Fraction} \\
\colhead{} & \colhead{(ppm)} & \colhead{I} & \colhead{II} 
& \colhead{III} & \colhead{IV}}
\startdata
H  & 1.00E+06 &    0.766 &    0.234&      0.00 &     0.00  \\
He & 1.00E+05 &    0.538 &    0.448&    0.0135 &     0.00  \\
C  &     309. & 0.000407 &    0.971&    0.0287 &     0.00  \\
N  &     49.0 &    0.650 &    0.350&  0.000141 &     0.00  \\
O  &     389. &    0.782 &    0.218&  7.93E-05 &     0.00  \\
Ne &     123. &    0.115 &    0.618&     0.268 & 8.14E-06  \\
Na &     2.04 &  0.00146 &    0.842&     0.156 & 3.48E-06  \\
Mg &     5.37 &  0.00211 &    0.786&     0.212 &     0.00  \\
Al &   0.0794 & 7.69E-05 &    0.973&    0.0179 &  0.00940  \\
Si &     6.61 & 4.98E-05 &    0.997&   0.00239 & 2.84E-05  \\
P  &    0.219 & 0.000174 &    0.977&    0.0227 & 9.94E-05  \\
S  &     14.5 & 8.44E-05 &    0.960&    0.0396 & 3.42E-06  \\
Ar &     2.82\tablenotemark{b} &    0.240 &    0.473&     0.287 & 5.08E-06  \\
Ca & 0.000407 & 2.71E-05 &   0.0164&     0.983 & 0.000230  \\
Fe &     2.04 & 0.000206 &    0.974&    0.0255 & 6.83E-06  \\
\enddata
\tablenotetext{a}{Model 18 from \citet{SlavinFrisch:2002}.}
\tablenotetext{b}{This is the Ar abundance used in Paper II.  
In this paper we use Ar/H=3.16 PPM from \citet{Holmgrenetal:1990}.}
\end{deluxetable}

\clearpage

\end{document}